\documentclass[prx,superscriptaddress,amsmath,amssymb,floatfix,twocolumn,showpacs,amsfonts,longbibliography,prl]{revtex4-1}
\usepackage{times}
\usepackage[varg]{txfonts}
\usepackage{textcomp}
\usepackage{graphicx}
\usepackage{subfigure}
\usepackage{tabu}
\usepackage{color}
\usepackage[plainpages=false,pdfpagelabels,colorlinks=true,linkcolor=red,urlcolor=blue,citecolor=blue,pdftitle={Title},pdfauthor={},pdfdisplaydoctitle=true,pdfduplex=DuplexFlipLongEdge]{hyperref}
\usepackage{braket}
\usepackage{float}
\usepackage{overpic}
\usepackage{gensymb}
\usepackage{mathrsfs}
\usepackage{tikz}
\usepackage{bm}
\usepackage{multirow}
\usepackage{amsfonts}
\usepackage{amsmath}
\usepackage{mathrsfs}
\usepackage{amssymb}
\usepackage{paralist}
\usepackage{indentfirst}
\usepackage{ulem}
\usepackage{soul}
\usepackage{cancel}
\usepackage{CJK}
\usepackage{ctable}

\newcommand{\brak}[2]{\langle{#1}|{#2}\rangle}

\newcommand{\process}[1]{\textcircled{\footnotesize{#1}}}
\allowdisplaybreaks

\hypersetup{
           breaklinks=true,   
           colorlinks=true,   
           pdfusetitle=true,  
        }
        
\newcommand{\tcr}{\textcolor{red}}

\begin{document}
\title{Spinon Singlet Pairing: Origin of $d$-Wave Sign Structure in a Partially-Filled Stripe}
\author{Jia-Long Wang}
\affiliation{Beijing Computational Science Research Center, Beijing 100193, People's Republic of China}
\author{Shi-Jie Hu}
\email{Corresponding author: shijiehu@csrc.ac.cn}
\affiliation{Beijing Computational Science Research Center, Beijing 100193, People's Republic of China}
\affiliation{Department of Physics, Beijing Normal University, Beijing 100875, People's Republic of China}
\author{Xue-Feng Zhang}
\email{Corresponding author: zhangxf@cqu.edu.cn}
\affiliation{Center of Quantum Materials and Devices, Chongqing University, Chongqing 401331, China}
\affiliation{Department of Physics and Chongqing Key Laboratory for Strongly Coupled Physics, Chongqing University, Chongqing 401331, China}

\begin{abstract}
Significant research advances have led to a consensus that the Fermi-Hubbard model and its extended variants are archetypical frameworks for elucidating the intertwined relationship between stripe orders and superconductivity in hole-doped high-$T_c$ materials. Notably, the Hubbard quantum simulator has recently achieved several remarkable breakthroughs, e.g., being cooled down to the cryogenic regime and enabling the observation of stable fluctuating stripes. However, the microscopic mechanism underlying the $d$-wave pairing of electrons in the presence of stripes at low temperatures remains poorly understood due to the intricate interplay between strongly correlated effects and non-negligible thermal fluctuations. Here, we conduct a close investigation of a partially-filled stripe in the representative $t$-$J$ and $t$-$t'$-$U$ models with both numerical and analytical methods. Analogous to quantum gas microscopy, the perfect sampling technique allows us to obtain the high-confidence-level statistics of the Fock basis states appearing in the ground-state wavefunction. In a novel physical paradigm, these data demonstrate that two spinons with opposite chiralities tend to pair spontaneously into a singlet state, thereby naturally giving rise to the $d$-wave pairing pattern. Then, using the updated effective theory of quantum colored string, we have reconstructed the wavefunction and have determined the nature of spinon pairing and its connection to the $d$-wave sign structure of pair-pair correlation. Furthermore, spinon singlet pairs enable the establishment of a long-range pair-pair correlation between the two stripes.
Our work offers new insights into the microscopic physics of stripes and paves the way for further exploration of multi-stripe-mediated pairing mechanisms in the Fermi-Hubbard model.
\end{abstract}
\maketitle

\textbf{\textit{Introduction}}.--Establishing the correct microscopic mechanisms for the pairing process is a crucial step in understanding high-$T_c$ superconductivity~\cite{BCS, hightc_review0, hightc_review1, hightc_review2, hightc_review3, hightc_review4}. In hole-doped cuprates, experiments have provided strong support for the idea of local electron pairing in the widely observed $d$-wave superconducting states~\cite{paring_exp0, paring_exp1, paring_exp_yayu}.
Numerical simulation results, which are consistent with experimental observations~\cite{Exp_stripe1, Exp_stripe2}, have also suggested an intertwined relationship between the stripe phase and superconductivity~\cite{stripe_review1, stripe_review2}, extending beyond the Hubbard model~\cite{Hubbard_Science, Hubbard_PRL, Hubbard_PRX, Hubbard_shiwei, Hubbard_yuanyao}, to include a wide range of models, e.g., $t$-$J$, $t$-$t’$-$J$~\cite{White_stripe, t-J1, t-J_ext1, t-J_ext2, t-J_ext3, ttj_science, tj_jiang, ttj_liwei}. More recently, simulations of the Hubbard model with next-nearest-neighbor (NNN) hopping $t’$ terms using density matrix renormalization group (DMRG) and quantum Monte Carlo (QMC) methods have revealed the coexistence of $d$-wave superconductivity and partially-filled stripes~\cite{Hubbard_shiwei}. Meanwhile, the origin of the pseudogap appears to be closely related to the melting of the stripe phase that exhibits pre-pairing~\cite{paring_pg,pg_science}.
Therefore, it is widely accepted that the single-band $t$-$t’$-$U$ model and the $t$-$J$ model can capture the essential physics of high-$T_c$ superconductivity. 

\begin{figure}[t!]
\centering
\includegraphics[width=0.99\linewidth]{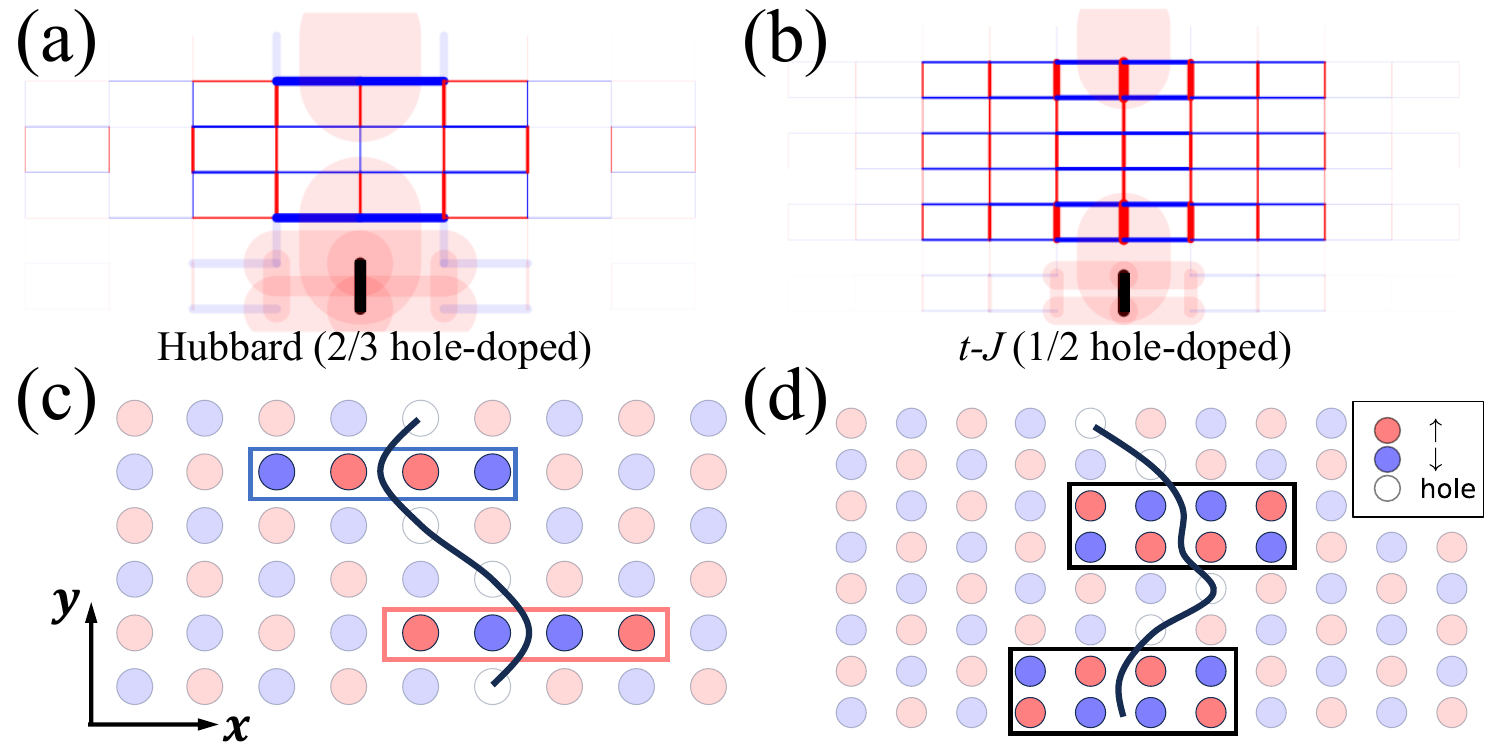}
\caption{The PPC function, starting from the target $y$-bond (black line with a width corresponding to a magnitude of $0.01$), is calculated with the DMRG method for (a) the Hubbard model including the NNN hopping $t'=-0.2$ with $U=12$ and $\mathbf{L}=(L_x,\, L_y)= (9,\, 6)$. (b) the $t$-$J$ model at $J=0.6$ and $\mathbf{L}= (11,\, 8)$.
The red (blue) ribbons represent positive (negative) PPC values.
The largest-weight snapshots are obtained by using perfect sampling the ground-state wavefunction for (c) the $t$-$t'$-$U$ model and (d) the $t$-$J$ model.
The spinons $\uparrow\downarrow\downarrow\uparrow$ ($\downarrow\uparrow\uparrow\downarrow$) with chirality $\Uparrow$ ($\Downarrow$) are highlighted in red (blue) rectangles, and their pairing is marked with a black frame.
}\label{fig01}
\end{figure}

Unveiling the mysteries of the Fermi-Hubbard model is also one of the critical goals for quantum simulation~\cite{ol_string_pattern,QS_1,ol_stripe,ol_pan,ol_2025,ol_ye,QS_2,QS_3}. Quantum gas microscopy, as a hallmark method, provides an ideal tool for instantaneously capturing the distribution of fermions~\cite{ol_string_pattern, ol_stripe}, so microscopic dynamics can be observed directly. Such in situ detection is particularly vital for strongly correlated systems~\cite{ol_spincharge}, where the quasi-particle size is often comparable to the lattice spacing, e.g., a spinon, a Majorana fermion, or a fracton. Therefore, it is urgent to elucidate the microscopic mechanism of local pairing inside the stripe and its relationship to the $d$-wave sign structure of the pair-pair correlation (PPC) function.

In this letter, a partially-filled stripe on a square-lattice cylinder is systematically analyzed using both numerical and analytic methods. As demonstrated in Figs.~\ref{fig01}(a) and (b), which are obtained using large-scale DMRG simulations, the PPC function $G_{b,b'} = \braket{\Delta^\dag_b \Delta_{b'}^{\phantom{\dag}}}$ at bonds $b$ and $b'$ clearly shows the $d$-wave patterns around the stripe in both the $t$-$t'$-$U$ Hubbard and $t$-$J$ models. Specifically, $G_{b,b'}$ between a target $y$-bond and a distant $x$-bond exhibits a negative sign, as indicated by the blue ribbons. Then, with the help of the perfect sampling technique~\cite{perfect_sampling} and an updated effective theory of quantum colored string (QCS) model (QCSM) {that extends our earlier proposed QCSM framework~\cite{color_string}}, we find that the spinons (topological point defects)~\cite{tj_spinon} with spin configurations $\uparrow\downarrow\downarrow\uparrow$ and $\downarrow\uparrow\uparrow\downarrow$ prefer to pair in a singlet form along the QCS (topological line defect), as shown in Figs.~\ref{fig01}(c) and (d). These spinon singlets, which interact with quantum fluctuations along the stripe, directly constitute the $d$-wave pairing pattern of electrons. Therefore, our work quantitatively shows that the spinons confined within the stripe can spontaneously pair in singlet form, so that the $d$-wave sign structure of the PPC function can be built, which challenges the existing hole-pairing scenario~\cite{hole_pair_weng,hole_pair_fabian}.

\begin{figure}[t!]
\centering
\includegraphics[width=0.99\linewidth]{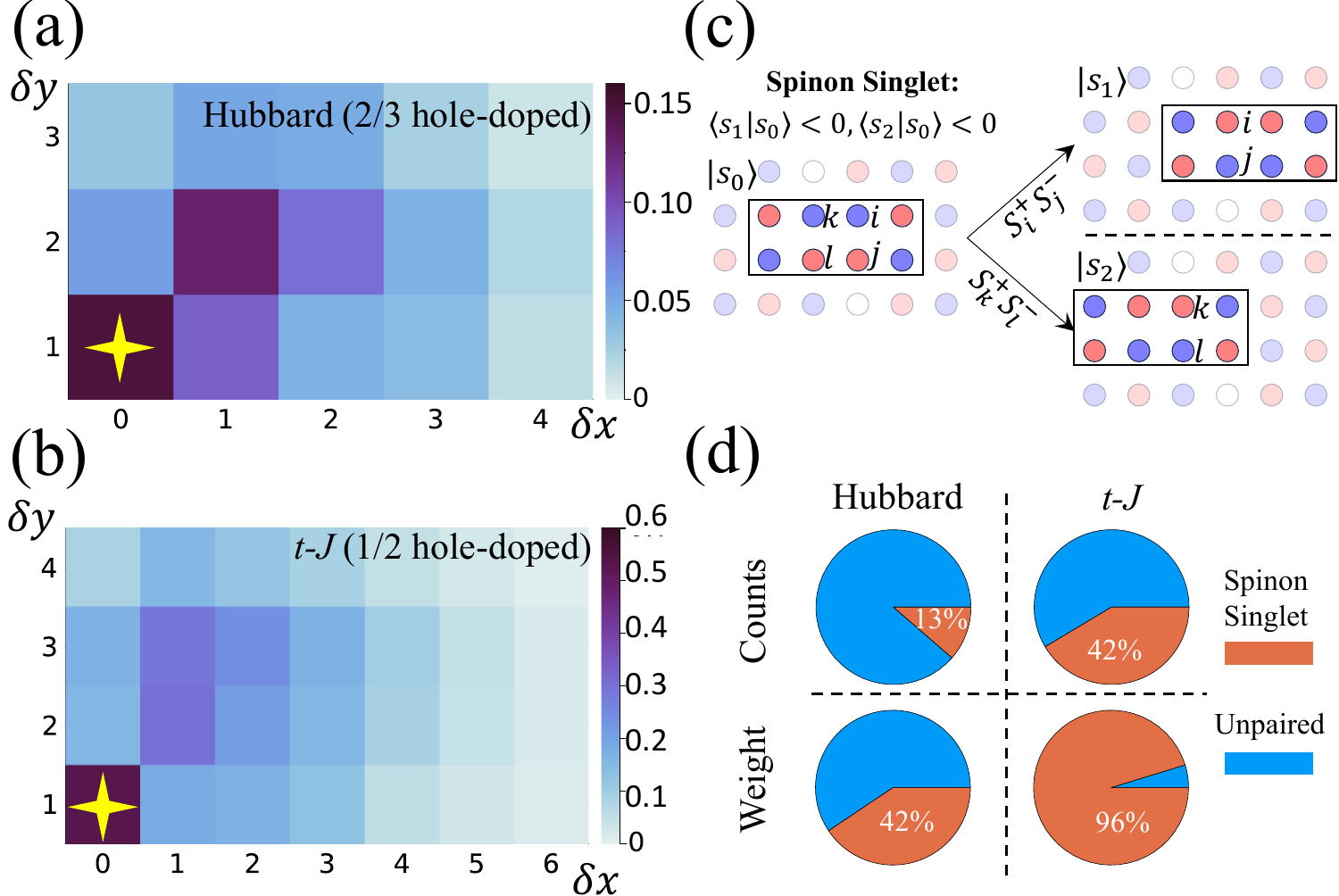}
\caption{(a,b) Spinon correlation function with spinon distances $(\delta x,\delta y)$ in the $x$ and $y$ directions.
(c) Two spin-flipping processes for identifying the ``spinon singlet". (d) The statistics of counts and weights of snapshots for different categories.
All the data are obtained via the perfect sampling technique, and the parameters are the same as those used in Fig.~\ref{fig01}.
}\label{fig02}
\end{figure}

\textbf{\textit{Models}}.--The $t$-$J$ model is described by the following Hamiltonian:
\begin{equation}
   \label{eq:tjmodel}
	    H =-t\sum_{\langle \ell,\ell'\rangle,\sigma}\left(f_{\ell,\sigma}^\dag f_{\ell',\sigma}^{\phantom{\dag}}+\textrm{h.c.}\right)+J\sum_{\langle \ell,\ell'\rangle} \left(\bold{S_\ell \cdot S_{\ell'}} - \frac{n^{\phantom{\dag}}_\ell n^{\phantom{\dag}}_{\ell'}}{4}  \right)\, ,
\end{equation}
where $f_{\ell,\sigma}^\dag$, $f_{\ell,\sigma}^{\phantom{\dag}}$ and $n_{\ell,\sigma}^{\phantom{\dag}} = f_{\ell,\sigma}^\dag f_{\ell,\sigma}^{\phantom{\dag}}$ represent the creation, annihilation, and particle number operators for an electron with spin $\sigma=\uparrow$,$\downarrow$ at site-$\ell$, respectively.
The term $\Delta_b = f_{j,\downarrow} f_{i,\uparrow} - f_{j,\uparrow} f_{i,\downarrow}$ describes the annihilation of a singlet pair at bond $b = (i,\ j)$, where site $i$ has a higher index than site $j$, according to an earlier site-labeling convention~\cite{color_string}.
The particle number operator at site $\ell$ is defined as $n_\ell = n_{\ell,\uparrow} + n_{\ell,\downarrow}$ with double occupation forbidden, while the spin components at site $\ell$ are given by $\bold{S_\ell}=(S^x_\ell,S^y_\ell,S^z_\ell)$.
The nearest-neighbor (NN) hopping amplitude $t=1$ defines the energy unit.
We set the spin interaction strength to $J$=0.6, which favors a $\pi$-phase-shifted stripe in the ground state.
For the $t$-$t'$-$U$ model~\cite{Hubbard_shiwei}, we set the on-site repulsion strength to $U=12$ and introduce an NNN hopping amplitude of $t'=-0.2$ to stabilize the partially-filled stripe.
{In the \textbf{\textit{Supplementary Material}} (\textbf{\textit{SM}}), additional benchmarks for $24$ other parameter sets are presented~\cite{SM}.}

\textbf{\textit{Spinon {Singlet} Pairing}}.--In quantum gas microscopy, the instantaneous configuration of the fermions can be photoed using the fluorescence excited states~\cite{ol_string_pattern}. Then, the numerous snapshots cannot only be used for quantum state tomography but also provide a visual physical picture for exploring the possible mechanism. A perfect sampling method can mimic such in situ measurements and produce snapshots by randomly projecting the matrix product state (MPS) to different Fock states $\ket{s}$. Beyond quantum gas microscopy, the coefficient $\alpha_s=\brak{s}{\psi_\text{D}}$ and the weight $|\alpha_s|^2$ can be extracted. In this work, five million snapshots are sampled for each parameter set, which has been verified to be sufficient to draw robust conclusions [see details in \textbf{\textit{End Matter {A}}}]. 

In Figs.~\ref{fig01}(c) and (d), the largest-weight snapshot of both the $t$-$t'$-$U$ Hubbard and $t$-$J$ models contains rich information. A clear $\pi$-phase shift is evident between two antiferromagnetic (AF) domains.
The domain wall is composed of two types of color particles (CPs): spinons $\uparrow\downarrow\downarrow\uparrow$ or $\downarrow\uparrow\uparrow\downarrow$ and holons $\uparrow\circ\downarrow$ or $\downarrow\circ\uparrow$. Including additional CPs: dual-holes $\uparrow\circ\circ\uparrow$ or $\downarrow\circ\circ\downarrow$ (discussed later), we can label them with color indices $c = \textbf{r}$, $\textbf{g}$, $\textbf{b}$. CPs can be further distinguished by their chiralities $\chi$, defined according to the orientation of the leftmost spin.
For example, for spinons, we have $\textbf{r}_\Uparrow=\uparrow\downarrow\downarrow\uparrow$ and $\textbf{r}_\Downarrow=\downarrow\uparrow\uparrow\downarrow$. Then, we can immediately find that the domain wall is a topological line defect, which is also named as QCS composed of CPs~\cite{color_string}. In Fig.~\ref{fig01}(d), four spinons spontaneously form two pairs with opposite chiralities reminiscent of singlet pairing.

In Figs.~\ref{fig02}(a) and (b), the snapshot statistics of spinon distances shows that two spinons most commonly appear at a distance $(\delta x,\delta y)=(0,1)$ in both the $t$-$t'$-$U$ Hubbard and $t$-$J$ models, which directly supports the formation of spinon pairs identified in Fig.~\ref{fig01}(d).
Interestingly, as depicted in Fig.~\ref{fig02}(c), the spin exchange operation on a $y$-bond can move the spinon pair along the $x$-axis while simultaneously reversing the chirality sequence.
Consequently, for a sample with a spinon pair $\ket{s_0}$, two states $\ket{s_1}$ and $\ket{s_2}$ can be generated by exchanging adjacent spins.
We find that for almost all spinon pairs, both overlaps $\brak{s_1}{s_0}$ and $\brak{s_2}{s_0}$ are negative, which marks the spinon pair as a ``spinon singlet''.
Sometimes, the singlet pair is destroyed due to the hole hopping and spin flipping [see {\textbf{\textit{SM}}}~\cite{SM}].
Samples lacking a spinon pair are categorized as ``unpaired" [Fig.~\ref{fig01}(c)].
The global snapshot statistics, shown in Fig.~\ref{fig02}(d), reveal that nearly all spinon pairs form singlets, as indicated by the negligible fraction of ``invalid" configurations.
Meanwhile, although the count of ``spinon singlet" is lower than that of the ``unpaired" part, the ratio of ``spinon singlet" to the total weight increases significantly.

Based on the perfect sampling data, we conclude that CPs tend to spontaneously organize into a stripe with a $\pi$-phase shift (i.e., a fluctuating QCS) in both the hole-doped $t$-$J$ and $t$-$t'$-$U$ models. The spinons prefer to pair in a singlet form.
Although quantum fluctuations can temporarily break the spinon singlet and ruin the local AF bonds, they can enhance the kinetic energy of the spinon singlet via higher-order processes.
Subsequently, we employ the {updated} effective theory of QCS~\cite{color_string} to uncover the connection between spinon singlets and the $d$-wave sign structure of the PPC function in the simplest case of the $t$-$J$ model.

\begin{figure}[t!]
\centering
\includegraphics[width=0.99\linewidth]{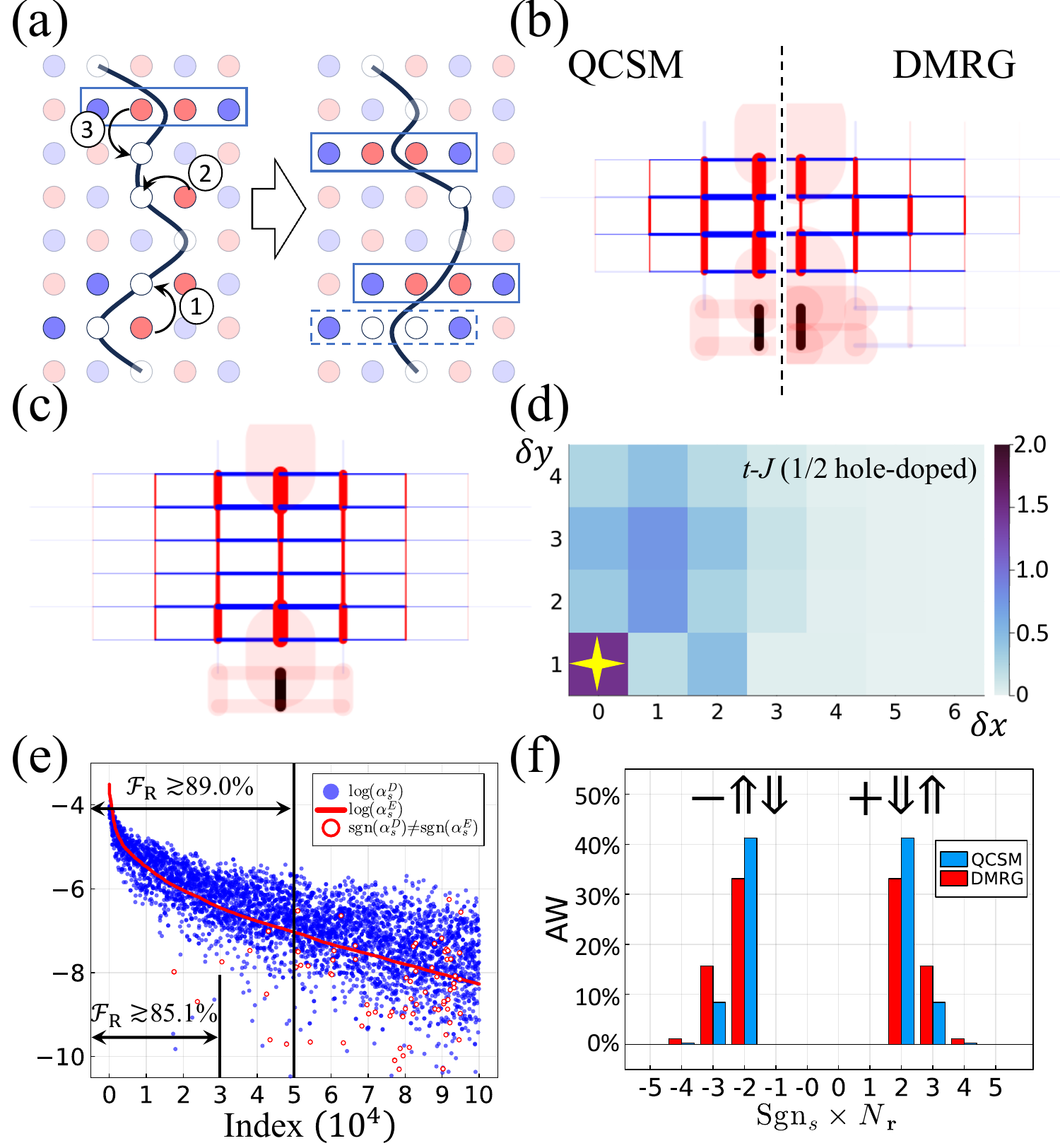}
\caption{(a) A schematic diagram shows the dynamics of QCS. The black arrows with numbers in circles label different dynamic processes. The dashed rectangle marks the dual-hole. (b) The PPC function $G_{b,b'}$ for a $2/3$ hole-doped stripe with $\mathbf{L} = (11,\, 6)$. (c) The PPC function $G_{b,b'}$ and (d) the spinon correlation function with different distances for a $1/2$ hole-doped stripe with $\mathbf{L} = (11,\, 8)$, calculated using the effective theory. (e) The amplitudes of expansion coefficients $\alpha_s^{\text{E}}$ and $\alpha_s^{\text{D}}$, and (f) distributions of AW in $\ket{\psi_\text{E}}$ and $\ket{\psi^\Omega_\text{D}}$ with $\mathbf{L}= (11,\, 6)$.
}\label{fig03}
\end{figure}

\textbf{\textit{Theory of QCS}}.--As demonstrated in Fig.~\ref{fig03}(a), the hopping process~\process{1} between two holons can create a spinon and a dual-hole, forming an \textbf{r}-\textbf{b} pair at adjacent rows.
The QCS theory imposes the constraint that $\sum_c n_c=1$ in each row.
Within this restricted Hilbert space $\Omega$, the stripe is described by a QCS. 
To account for the displacement of CPs along the $x$-axis between adjacent rows, an effective spin field $\Gamma^z\in\mathbb{Z}$ must be included. Then, the effective model for a single QCS can be explicitly derived [see Ref.~\cite {color_string}, \textbf{\textit{End Matter {B}}}, and \textbf{\textit{SM}~\cite{SM}} for details].

The AF background imposes an attractive diagonal interaction $\propto|\Gamma^z|$ between CPs, which acts as the tension energy of QCS.
In addition to the creation of an \textbf{r}-\textbf{b} pair and its reverse process, the electron hopping can deform the QCS and move the CPs, corresponding to processes \process{2} and \process{3} in Fig.~\ref{fig03}(a), respectively.
{In contrast to the previous QCSM derived from the $t$-$J_z$ model~\cite{color_string},} we have incorporated the portion of spin exchange interaction relevant to the spinons in the updated effective model.
{For example, spin exchange interactions in the $x$ and $y$ directions contribute to $x$-directional movement of single spinons and spinon pairs, respectively, and are added to the updated QCSM as additional terms $H^{(\text{ex1})}_\text{o}$ and $H^{(\text{ex2})}_\text{o}$.
The spin exchange term $H^{(\text{ex1})}_\text{o}$ provides the kinetic energy of spinons in the $x$ direction.
The most critical term $H^{(\text{ex2})}_\text{o}$ swaps the chiralities of spinons between adjacent rows, which is the key mechanism for spinon singlet pairing.}
The remaining part, which only disrupts the local order in AF domains, is argued to renormalize the parameters of the effective model without introducing additional QCSM terms.
In this work, only the spin exchange interaction {strength $J_\pm$} is adjusted to account for this renormalization effect [see \textbf{\textit{End Matter {B}}}].

Clearly, the potential energy suppresses the vibration of the QCS, while the quantum fluctuations introduced by the off-diagonal terms enhance such vibration. The interplay between the dynamics of CPs and local quantum fluctuations yields rich physics of the stripe.
The ground-state wavefunction $\ket{\psi_\text{E}} = \sum_{s \in\Omega} \alpha_s^{\text{E}} \ket{s}$ of the effective QCSM is obtained through exact diagonalization (ED) with truncation $|\Gamma^z|\le\Gamma^z_{\mathrm{max}}$, where $\alpha_s^{\text{E}}$ gives the expansion coefficient of the basis $\ket{s}$.
Then, the PPC function shown in Fig.~\ref{fig03}(b), with $\Gamma^z_{\mathrm{max}}=9$, can be obtained and both QCS theory and DMRG results present a clear and consistent $d$-wave pattern. Meanwhile, the PPC function and spinon correlation of a $1/2$ hole-doped stripe are also calculated using the effective QCSM and plotted in Figs.~\ref{fig03}(c) and (d). They present distributions similar to the DMRG results in Fig.~\ref{fig01}(b) and Fig.~\ref{fig02}(b), respectively. These results demonstrate that the effective theory of QCS can {semi-quantitatively} describe the physics of a partially-filled stripe in the $t$-$J$ model. Subsequently, we anticipate that the relationship between $d$-wave pairing and spinon singlets is encoded in the wavefunction. 

The QCS theory offers a unique advantage for analyzing the ground-state wavefunction and identifies the important Fock basis vectors contained within the space $\Omega$.
For $L_y=6$ and $\Gamma^z_{\mathrm{max}}$=5 in the effective theory, the Hilbert space dimension reaches $286,098$, which is about $0.25\%$ of the dimension used in DMRG calculations with $M=16,384$. After projecting the wavefunction $\ket{\psi_\text{D}}$ onto $\Omega$, we obtain the wavefunction $\ket{\psi_\text{D}^\Omega} = \sum_{s \in \Omega} \alpha^\text{D}_s \ket{s}$, where $\alpha^\text{D}_s = \braket{s | \psi_\text{D}}$.
We then calculate the renormalized fidelity $\mathcal{F}_\text{R} = \lvert \braket{\psi_\mathrm{E} | \psi_\mathrm{D}^\Omega} \rvert / \sqrt{W}$, with $W = \braket{\psi_\text{D}^\Omega | \psi_\text{D}^\Omega}$.
Using $M=16,384$ in DMRG, we find $\mathcal{F}_\text{R} \approx 91.6\%$. As shown in Fig.~\ref{fig03}(e) arranged in descending order of the amplitude $|\alpha_s^{\text{E}}|$, first $50,000$ bases can provide very high fidelity, and the signs of $\alpha_s^\text{E}$ and $\alpha_s^\text{D}$ differ in only a few cases, impressively.

The basis $\ket{s}$ can be characterized by the number of spinons $N_\textbf{r}$ and the sign $\text{Sgn}_s \equiv \text{sgn}(\alpha_s^{\text{E}})$.
As shown in Fig.~\ref{fig03}(f), the two-spinon bases with $N_\textbf{r}$=2 dominate the ground-state wavefunction, with the accumulated weights (AW) being symmetric with respect to $\text{Sgn}_s$=$\pm 1$.
The distribution of three-spinon bases exhibits similar features, though with much lower weights, allowing us to neglect the contributions from other bases with $N_\textbf{r}>3$.
Given that the lower and upper spinons in the two-spinon bases carry chiralities $\chi_1$ and $\chi_2$, we only need to consider configurations where $\chi_1 \ne \chi_2$, as the ground state lacks net magnetization.
Interestingly, we find that $\text{Sgn}_s$ for a two-spinon basis depends on the order of the chiralities, as highlighted in Fig.~\ref{fig03}(f) (with only $\approx4\%$ mismatches).
Specifically, $\text{Sgn}_s = \mp 1$ for $\chi_1\chi_2 = \Uparrow \Downarrow$ and $\Downarrow \Uparrow$, respectively, which exactly matches the sign structure of the spinon singlet. 
The three-spinon bases $\ket{s}$ arise from local vacuum fluctuations of the two-spinon bases $\ket{s'}$.
If the emergent dual-hole is far from the two spinons in $\ket{s}$, the sign of its expansion coefficient is simply given by $\text{Sgn}_s = \text{Sgn}_{s'} \text{Sgn}_0$.
Here, the sign $\text{Sgn}_0$ depends on the specifics of electron hopping in the process from $\textbf{g}\text{-}\textbf{g}$ to $\textbf{r}\text{-}\textbf{b}$~\cite{color_string}. Considering that the wavefunction $\ket{\psi_\text{E}}$ exhibits a robust sign structure of spinon singlets, a long-range $d$-wave pattern and a high fidelity with $\ket{\psi_\text{D}}$, we expect that the link between spinon singlets and $d$-wave sign pattern can be effectively diagnosed through detailed analysis of the wavefunction.

\begin{figure}[t!]
\centering
\includegraphics[width=0.99\linewidth]{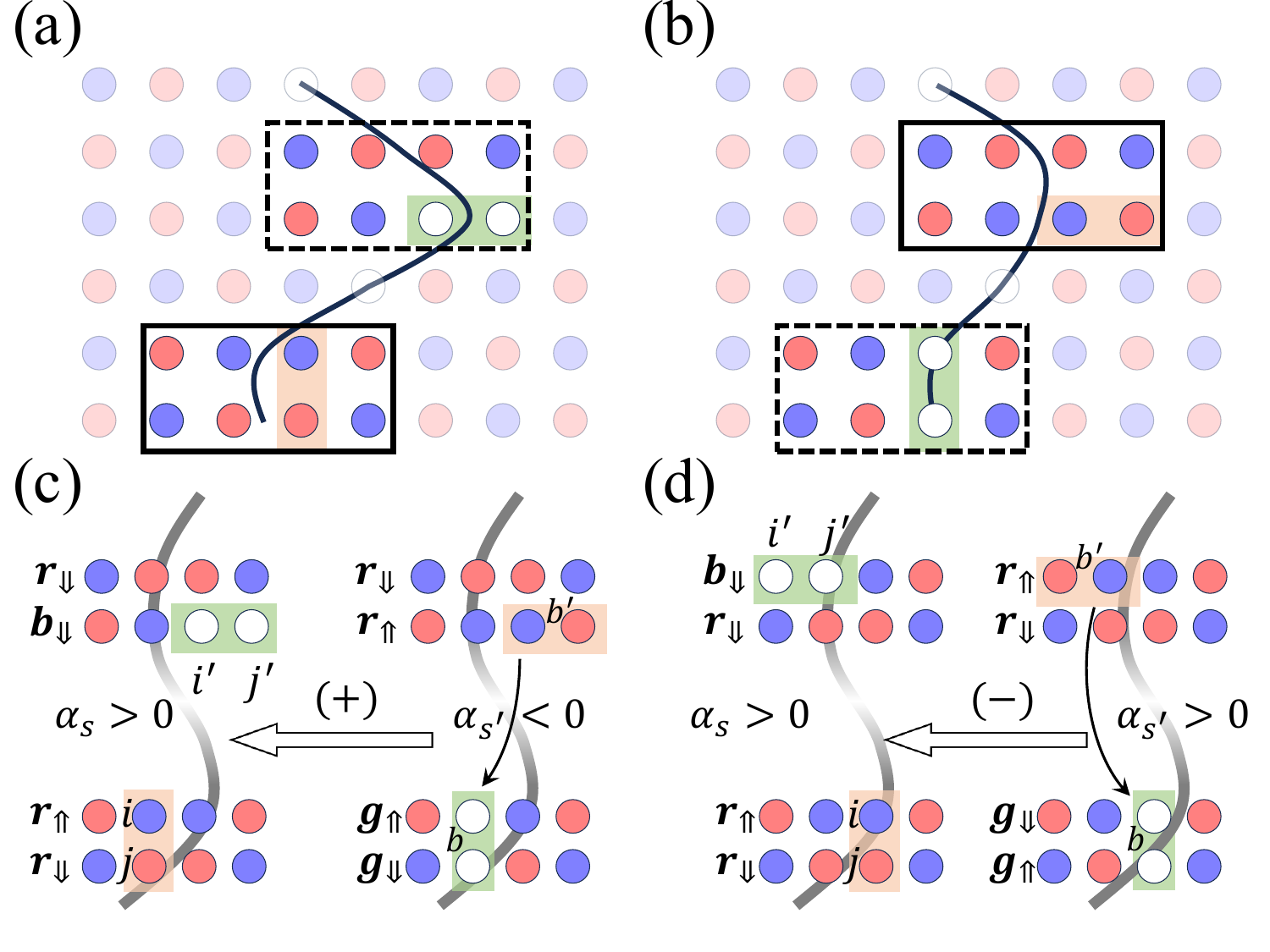}
\caption{Schematic diagrams illustrate a pair of (a) three-spinon and (b) two-spinon bases, which most significantly contribute to $G_{b,b'}$ for an $x$-bond $b'$ and a $y$-bond $b$ in a $2/3$ hole-doped stripe with $\mathbf{L}=(11,\, 6)$.
The dashed rectangles enclose a dule-hole-spinon in (a) and a hole-hole pair in (b), respectively.
The processes of moving an electron pair from $x$-bond $b'$ to $y$-bond $b$ (c) with and (d) without changing the chirality sequence are shown.
The sign of $\text{Sgn}_\Delta = \pm$ is indicated above the black hollow arrows, corresponding to cases (c) without and (d) with spin exchange, respectively. The red (green) rectangle highlights the electron (hole) pair. 
}\label{fig04}
\end{figure}

\textbf{\textit{$d$-wave sign structure}}.--At leading order, the PPC function receives a contribution $G^{s,s'}_{b,b'} = \alpha^*_s \alpha^{\phantom{\dag}}_{s'} \braket{s | \Delta^\dag_b \Delta^{\phantom{\dag}}_{b'} | s'}$ from the basis $\ket{s}$ and $\ket{s'}$.
At long distances, the sign of $G^{s,s'}_{b,b'}$ is given by $\text{Sgn}\left(G^{s,s'}_{b,b'}\right) = \text{Sgn}_s \text{Sgn}_{s'} \text{Sgn}_\Delta$, where $\text{Sgn}_\Delta$ accounts for spin exchange between $\Delta_b$ and $\Delta_{b'}$. Then, the PPC function between two $y$-bonds can be directly understood. The operator $\Delta_{b'}$ can annihilate two electrons with opposite spins at $y$-bond $b'$ in a two-spinon basis $\ket{s'}$, and then $\Delta^\dag_{b}$ fills two holes at $y$-bond $b$ to generate the basis $\ket{s}$. If the chirality sequence $\chi_1\chi_2$ in $\ket{s}$ and $\ket{s'}$ remains unchanged, then $G^{s,s'}_{b,b'} > 0$.
However, if the chirality sequence is altered, the sign $\text{Sgn}_\Delta = -1$ must be included to compensate for the sign change caused by the altered chirality sequence.
As a result, $G_{b,b'} > 0$ always holds for two $y$-bonds at long distances, owing to the spinon singlet state. Similarly, $G_{b,b'}$ is positive for two $x$-bonds, which is related to the presence of a spinon at $x$-bond $b'$ and a dual-hole at $x$-bond $b$ [see \textbf{\textit{SM}}~\cite{SM} for details].

The most critical feature of the $d$-wave pattern is the negative PPC function $G_{b,b'}$ between a $y$-bond $b$ and an $x$-bond $b'$.
According to the effective theory, the nonzero contribution $G^{s,s'}_{b,b'}$ stems from a three-spinon basis $\ket{s}$ and a two-spinon basis $\ket{s'}$ at leading order.
The number of these basis pairs \{$\ket{s'}$,\ $\ket{s}$\} that contribute nonzero $G^{s,s'}_{b,b'}$ is quite limited, especially for bonds at long distances.
This allows us to systematically identify all contributing pairs and to uncover the underlying microscopic pairing mechanism. 

As demonstrated in Fig.~\ref{fig04}(a), in the three-spinon basis $\ket{s}$ of the largest-weight basis pair, the spinon pair provides two electrons at $y$-bond $b$, while the \textbf{r}-\textbf{b} pair provides two holes at $x$-bond $b'$.
Then, evaluating $G^{s,s'}_{b,b'}$, two bonds transform into two holons and a spinon pair with different chirality sequence on the two-spinon basis $\ket{s'}$ shown in Fig.~\ref{fig04}(b). Therefore, as analyzed in Fig.~\ref{fig04}(c), if the chirality sequence of the spinon pair is different between $\ket{s'}$ and $\ket{s}$, the $\text{Sgn}_\Delta$ has to be $+1$. By analogy, the same chirality sequence requires $\text{Sgn}_\Delta=-1$ [Fig.~\ref{fig04}(d)]. This analysis is also applicable to $G_{b,b'} < 0$ between an $x$-bond $b$ and a $y$-bond $b'$ at long distances [see more details in \textbf{\textit{SM}}{~\cite{SM}}].

\begin{figure}[t!]
\centering
\includegraphics[width=0.99\linewidth]{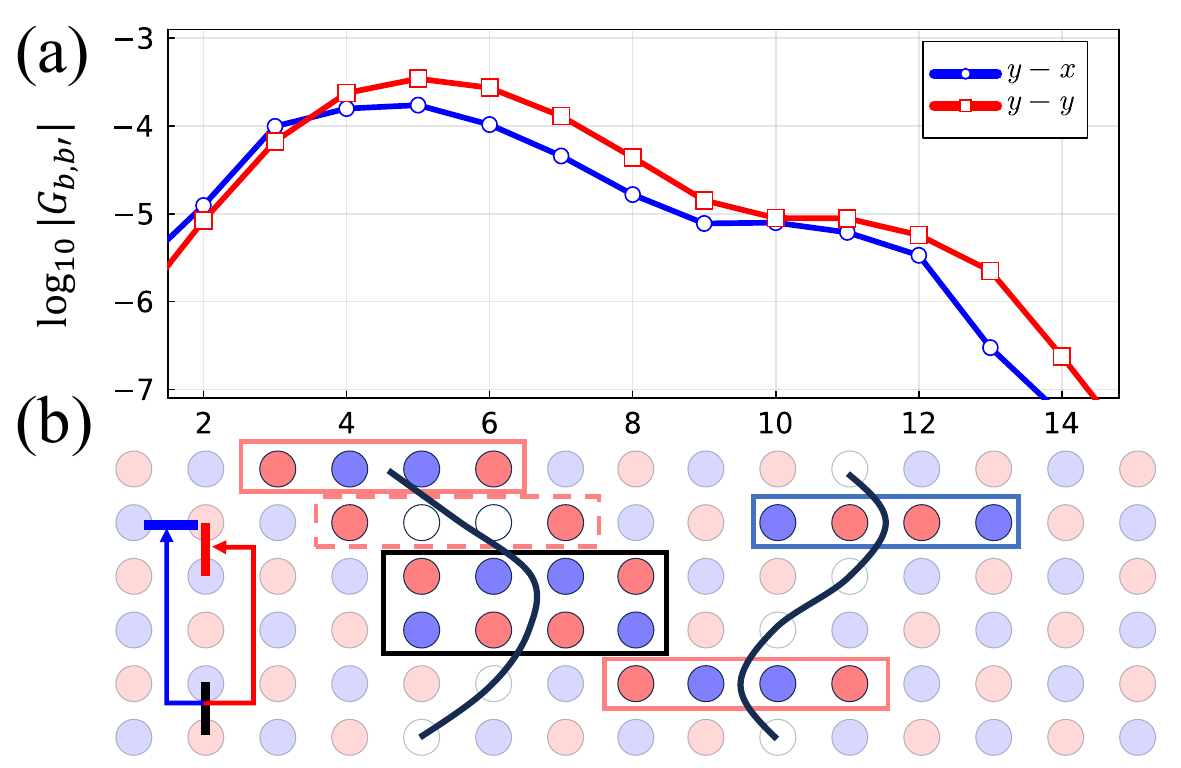}
\caption{(a) The PPC function $G_{b,b'}$ and (b) the largest-weight snapshot in the $t$-$J$ model with $N_h=8$, i.e., two $2/3$ hole-doped stripes, and $\mathbf{L}=(15,\, 6)$, calculated using the DMRG method. The $y$-$y$ ($y$-$x$) PPC function plotted in (a) corresponds to the correlation between the black target bond and the red $y$ (blue $x$) bonds in the same row.
}\label{fig06}
\end{figure}

\textbf{\textit{Conclusion and Discussion}}.--In both the $t$-$t'$-$U$ Hubbard and $t$-$J$ models, we identify a clear microscopic mechanism in which the spinon singlet, arising from a fluctuating two-spinon quantum colored string, leads to the emergence of {the} $d$-wave {sign structure of the pair-pair correlation function}. Building on the quantum colored string scenario we originally proposed~\cite{color_string} {and updated here}, we have developed a bottom-up effective theory to describe the correlations between distant spinon pairs in the ground state of a partially-filled stripe.
{We further perform a controlled tuning validation test, and unambiguously confirm that the mechanism for spinon singlet pairing originates from the spin exchange interaction, specifically the $J_\pm$ term of $H^{(\text{ex2})}_\text{o}$ defined in Eq.~\eqref{eqex2} [see details in \textbf{\textit{End Matter C}}].}

We also extend the DMRG simulation to a larger system with two stripes.
In Fig.~\ref{fig06}(a), the PPC function, starting from the target $y$-bond near the left edge column, is enhanced by the left QCS and further boosted by the right QCS over long distances. From the largest-weight snapshot displayed in Fig.~\ref{fig06}(b), we can clearly identify two QCSs and complex interplay between CPs, {which can be taken as a ``generalized spin-charge separation"}. It appears to depict microscopic harmonic dance of spinons: a spinon singlet that is activated within a QCS and tunnels into a neighboring QCS.
Therefore, we are confident in conjecturing that the spinon pair may also serve as an elementary quasi-particle in the multi-stripe phase of Fermi-Hubbard-like models. Meanwhile, as discussed in \textbf{\textit{End Matter {B}}}, the spinon singlet pairs have not only the lowest diagonal energy $H_d$, but they can also enlarge their pairing energy gap with the help of the spin exchange interaction. Definitely, the detailed microscopic mechanism in multiple stripes requires more future exploration.

Finally, we discuss possible connections to quantum simulations. Because of the rapid development of the quantum simulation using ultra-cold atoms~\cite{QS_1,QS_2,QS_3}, the local pairing of the spinons is expected to be clearly observed in optical lattices via quantum gas microscopy~\cite{QS_3}. In addition, the distribution of the PPC function can also be checked in the mixD system that simulates the extended $t$-$J$ model~\cite{mixD_1,mixD_2,mixD_3}. 

{\textbf{\textit{Note Added.}}---Recent applications of the updated QCSM quantitatively reproduce both the characteristic $4a_0 \times 4a_0$ plaquette structures and the robust $2a_0$ shift between positive- and negative-bias local density of states patterns observed in scanning tunneling microscopy of underdoped cuprates~\cite{stm_liang}. This quantitative agreement with key experimental observables thus provides independent experimental validation for the spinon singlet pairing scenario identified in this work.}

\section*{acknowledgment}
We would like to thank Tao Li, Hao Ding, and Yuan-Yao He for many helpful discussions, and acknowledge the ad hoc editor of PRX for our previous paper~\cite{color_string} who suggested the perfect sampling method. S. J. H. acknowledges funding from MOST Grant No.~2022YFA1402700, NSFC No.~U2230402, NSFC Grant No.~12174020. X.-F. Z. acknowledges funding from the National Science Foundation of China under Grants  No.~12274046 and No.~12547101, and Xiaomi Foundation / Xiaomi Young Talents Program.

\bibliography{ref}

\title{End Matter}

\maketitle

\renewcommand{\thefigure}{E\arabic{figure}}
\setcounter{figure}{0}

\renewcommand{\theequation}{M\arabic{equation}}
\setcounter{equation}{0}

\section*{End Matter}
\textit{{End Matter A:} \textbf{DMRG}}.--For consistency in the following discussions, we always choose odd values for $L_x$. While selecting other values of $L_x$, or using irregularly sheared open boundaries may introduce certain phase factors to spinon singlets, which do not alter conclusions regarding $d$-wave pairing.
To prevent the formation of a frozen stripe, we double the value of $J$ along the bonds connecting sites in two edge columns~\cite{color_string}.
In both the $t$-$J$ and $t$-$t'$-$U$ models, we have performed benchmarks of three key quantities (i.e., truncation errors, the PPC function, the sampling data, {see details in \textbf{\textit{SM}}}).

\textit{{End Matter B:} \textbf{Effective Hamiltonian of a QCS}}.--When multiple holes are introduced into the AF background, they spontaneously organize into an individual stripe with a $\pi$-phase shift. This one-dimensional, curved structure is inherently quantum in nature due to electron hopping, distinguishing it from its classical counterpart. As a result, it is referred to as a ``quantum string''~\cite{string_peter, dqcp01, zhou01, wanyuan, supersolid}. Building on our recent work~\cite{color_string}, an individual stripe can be described by a QCS composed of three types of CPs. By imposing the constraint that only one CP is allowed per row, the restricted Hilbert space can be constructed, from which the effective Hamiltonian can be derived.

\begin{figure}[b]
	\centering
	\includegraphics[width=0.99\linewidth]{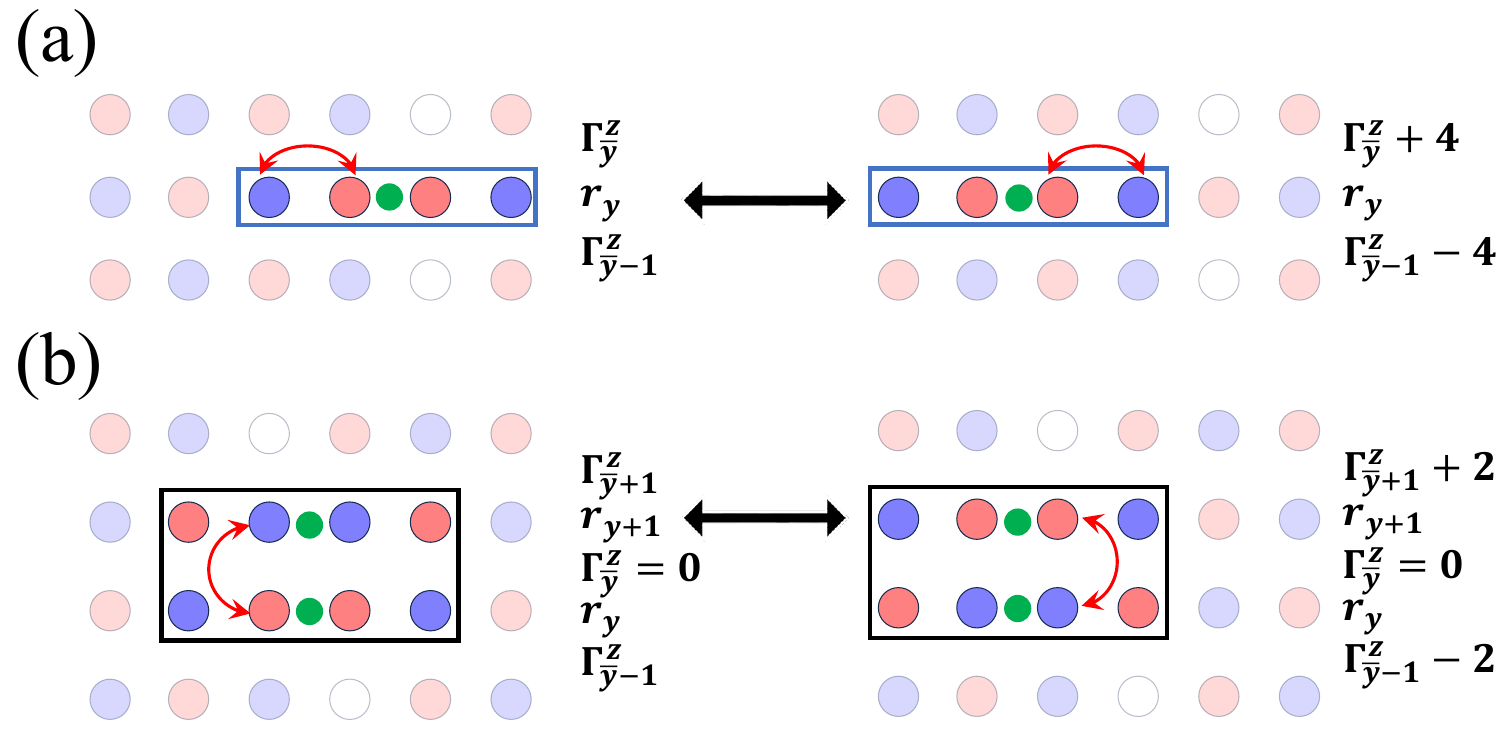}
	\caption{{Schematic images illustrate off-diagonal interactions $H^{(\text{ex1})}_\text{o}$ and $H^{(\text{ex2})}_\text{o}$ related to the spin exchange interaction in the (a) $x$ direction and (b) $y$ direction.}
	}\label{fige01}
\end{figure}
The $t$-$J$ model can be extended to a $t$-$J$-$\alpha$ model, which is explicitly written as
\begin{eqnarray}\label{eq:extendedtjmodel}
\begin{split}
H =&-t\sum_{\langle \ell,\ell'\rangle,\sigma}\left(f_{\ell,\sigma}^\dag f_{\ell',\sigma}^{\phantom{\dag}}+\textrm{h.c.}\right)\\
&+J\sum_{\langle \ell,\ell'\rangle} \left[ \alpha \left(S_\ell^x S_{\ell'}^x + S_\ell^y S_{\ell'}^y \right) + S_\ell^z S_{\ell'}^z - \frac{1}{4} n_\ell^{\phantom{\dag}} n_{\ell'}^{\phantom{\dag}} \right]\, ,
\end{split}
\end{eqnarray}
where the anisotropy parameter $\alpha$ is tunable.
To derive the effective Hamiltonian of QCS, we introduce the effective spin field $\Gamma^z_{\bar{y}} \in \mathbb{Z}$, where $\bar{y}$ labels the dual row in between rows $y$ and $y+1$. The corresponding raising (lowering) operators of the effective spin field and CPs are denoted as $\Gamma^+_{\bar{y}}$ ($\Gamma^-_{\bar{y}}$) and $c_{\chi}^\dag$($c_{\chi}$), respectively. Without loss of generality, we choose the position of the CP in row $y=1$ to label the position of QCS, denoted as $\mathcal{X}^{(\text{CS})}$. Consequently, an electron configuration has a QCS representation: $\ket{s} = \ket{\mathcal{X}^{(\text{CS})}, \{\Gamma^z\}, \{c_\chi\}}$, where the sequences $\{\Gamma^z\}$ and $\{c_\chi\}$ represent the effective spin fields and CPs across all rows, respectively. 

Building on the $t$-$J$-$\alpha$ model while neglecting local spin fluctuations inside the two AF domains, the effective model can be derived as:\begin{eqnarray}\label{eq:effHam}
\begin{split}
H^\text{CS}_\text{e} &\!=\! H_\text{d} \!+\! H_\text{o}^{(\textbf{g})} \!+\! H^{(\textbf{r}\text{-}\textbf{b} / \textbf{g}\text{-}\textbf{g})}_\text{o} \!+\!H^{(\textbf{r}\text{-}\textbf{g}/\textbf{g}\text{-}\textbf{r})}_\text{o} \!+\! H^{(\textbf{b}\text{-}\textbf{g}/\textbf{g}\text{-}\textbf{b})}_\text{o}\\
&+H^{(\text{ex1})}_\text{o}+H^{(\text{ex2})}_\text{o}.
\end{split}
\end{eqnarray}
The diagonal term $H_\text{d}$ describes the tension energy.
Off-diagonal terms describe various interaction processes between CPs: $H_\text{o}^{(\textbf{g})}$ accounts for electron hopping along the $x$-axis, while $H^{(\textbf{r}\text{-}\textbf{b} / \textbf{g}\text{-}\textbf{g})}_\text{o}$ captures the creation of an \textbf{r}-\textbf{b} pair and its reverse process, and the terms $H^{(\textbf{r}\text{-}\textbf{g}/\textbf{g}\text{-}\textbf{r})}_\text{o}$ ($H^{(\textbf{b}\text{-}\textbf{g}/\textbf{g}\text{-}\textbf{b})}_\text{o}$) describe the movement of spinons (dual-holes). In addition to these terms derived in details from the $t$-$J_z$ model~\cite{color_string}, the spin exchange interactions in the $x$ and $y$ directions contribute to the movement of the spinon and spinon pair in the $x$ direction, respectively, as demonstrated in Fig.~\ref{fige01}. Then, corresponding effective terms $H^{(\text{ex1})}_\text{o}$ and $H^{(\text{ex2})}_\text{o}$ only change the effective spin fields and can be written as:
\begin{eqnarray}
H^{(\text{ex1})}_\text{o}
\!=\! J_{\pm}\sum_{y} n^{(\textbf{r})}_y\left[(\Gamma_{\bar{y}}^+)^4(\Gamma_{\bar{y}-1}^-)^4\!+\!\textrm{H.c.}\right]\ ,
\end{eqnarray}
and
\begin{eqnarray}
H^{(\text{ex2})}_\text{o}\!=\!J_{\pm}\!\sum_{y} n^{(\textbf{r})}_{y+1}n^{(\textbf{r})}_y\left[(\Gamma_{\bar{y}+1}^+)^2(\Gamma_{\bar{y}-1}^-)^2\!+\!\textrm{H.c.}\right]\label{eqex2}
\end{eqnarray}
for $\Gamma_{\bar{y}}^z=0$, where $J_\pm=\alpha J/2$. 
The spin exchange term $H^{(\text{ex1})}_\text{o}$ 
provides the kinetic energy of spinons in the $x$ direction.
The other term $H^{(\text{ex2})}_\text{o}$ exchanges the chiralities of spinons located in two adjacent rows.
{As systematically demonstrated by the analysis presented in this work, adding the spin exchange terms $H^{(\text{ex1})}_\text{o}$ and $H^{(\text{ex2})}_\text{o}$ to the updated QCSM introduces spinon dynamics, a key feature absent in our previous QCSM that was originally derived from the $t$-$J_z$ model and solely accounts for electron dynamics~\cite{color_string}.
This modification substantially alters the physical behavior of the ground state.
Especially, $H^{(\text{ex2})}_\text{o}$ is indispensable for {binding spinons into singlet pairs as shown} in Fig.~\ref{fig02}(c).}

As mentioned before, the spin exchange interaction within the AF domains benefits the formation of spinon singlet, thereby lowering the ground-state energy.
However, these local quantum fluctuations introduce significant noise into the statistical data of spinons, which can overcurtain the characteristic signals of spinons in a QCS.
To prepare the ground-state wavefunction without this noise for the perfect sampling procedure outlined in this work, we apply an $z$-axis external staggered magnetic field to the two edge columns, thereby inducing a spin-rotation symmetry breaking [see details in \textbf{\textit{SM}}{~\cite{SM}}].
In QCSM, as described above, we focus solely on the spin exchange interaction relevant to spinons in QCSs and adjust the strength of this interaction to include additional renormalization effects from the AF domains.
Following these procedures and setting $\alpha=2$ [doubled spin exchange interaction strength], we successfully reproduce a $d$-wave pattern for a $1/2$ hole-doped stripe, as shown in Fig.~\ref{fig03}(c), which closely resembles the DMRG-calculated pattern in Fig.~\ref{fig01}(b).
Meanwhile, the spinon correlation functions obtained from QCSM in Fig.~\ref{fig03}(d) and DMRG in Fig.~\ref{fig02}(b) exhibit a similar distribution.

\textit{{End Matter C:} \textbf{Controlled tuning validation test}}.--The mechanism for spinon singlet pairing indeed originates from the spin exchange interaction, specifically the $J_\pm$ term of $H^{(\text{ex2})}_\text{o}$ defined in Eq.~\eqref{eqex2}, which binds spinons into {singlet pairs}.
\begin{figure}[t]
	\centering
	\includegraphics[width=0.99\linewidth]{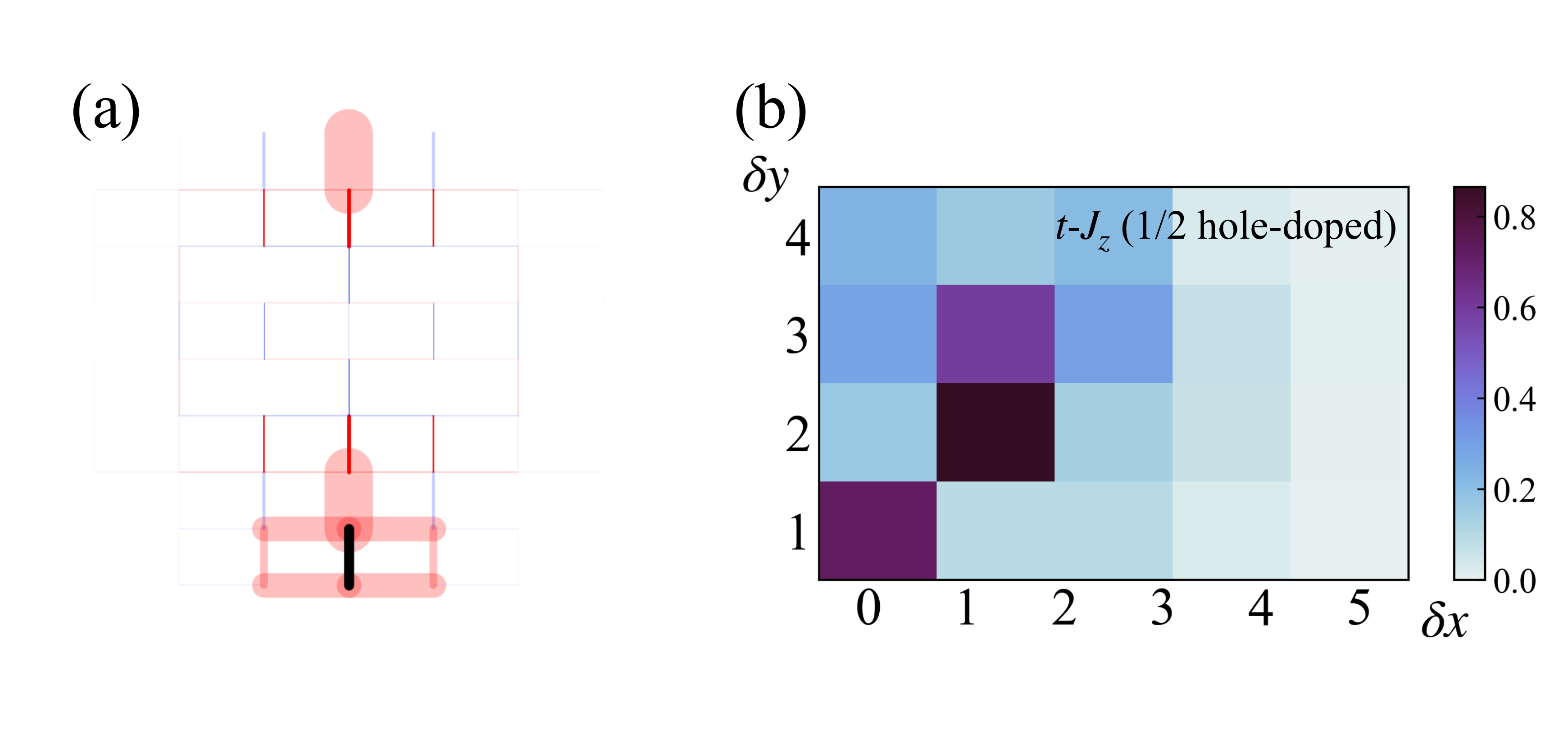}
	\caption{(a) The PPC function $G_{b,b'}$ for a $1/2$ hole-doped stripe with $\mathbf{L} = (11,\, 8)$. (b) Spinon correlation function with spinon distances $(\delta x,\delta y)$ in the $x$ and $y$ directions. Here, we target the $t$-$J_z$ model by setting $J_\pm = 0$.
	}\label{fige02}
\end{figure}
We investigate the pairing response of spinons when the proposed mechanism is removed. We achieve this by switching off the chirality exchange term, setting $J_\pm=0$ in the $t$-$J$ model at $\mathbf{L} = (11,\,8)$.
Following this adjustment, we observe the disappearance of $d$-wave pairing patterns at long distances as shown in Fig.~\ref{fige02}(a). 
Meanwhile, spinon pairs, characterized by $\delta x=0$ and $\delta y=1$, fail to dominate in Fig.~\ref{fige02}(b).
This demonstrates a causal relationship: without the spin exchange interaction, spinon singlet pairs do not form, and {the} $d$-wave {sign structure of the PPC function} vanishes.
Conversely, with $J_\pm \ne 0$, the energetic advantage drives singlet formation, and {the sign structure emerges} inevitably, as shown in Figs.~\ref{fig01}, \ref{fig02} and \ref{fig03} of the main text.
It is also consistent with a recent advancement in the explanation of experimental observations of the spatial local density of states~\cite{stm_liang}.

\newpage

\title{Supplementary Material}

\maketitle

\renewcommand{\thefigure}{S\arabic{figure}}
\setcounter{figure}{0}

\renewcommand{\theequation}{A\arabic{equation}}
\setcounter{equation}{0}

\section{PPC function}
\begin{figure}[h]
	\centering
	\includegraphics[width=0.8\linewidth]{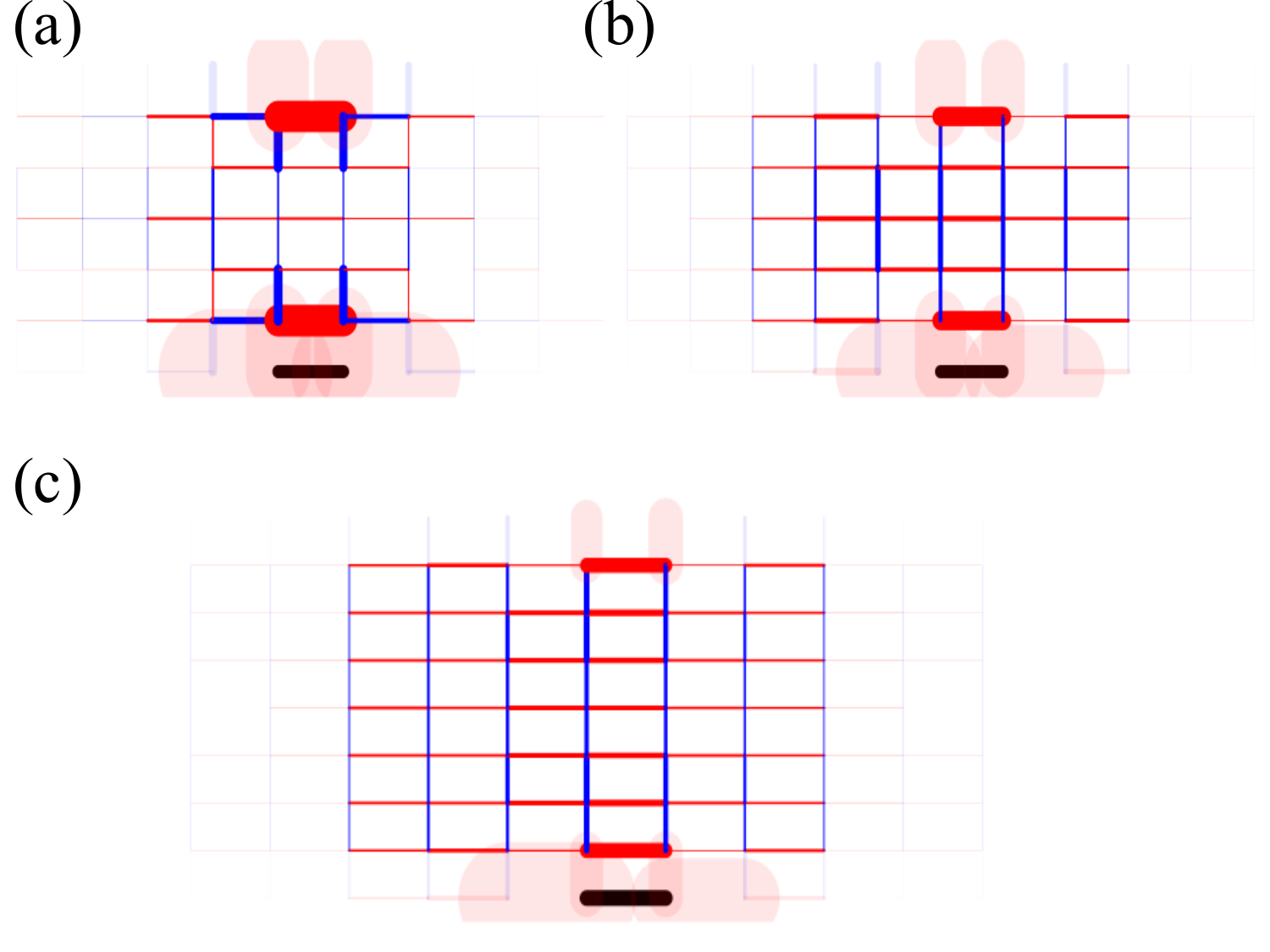}
	\caption{The PPC function, starting from a target $x$-bond (black line), is calculated using DMRG in the $t$-$t'$-$U$ and $t$-$J$ models. The parameters in (a-c) are the same as those used in Fig.~1(a), Fig.~3(b) and Fig.~1(b) of the main text, respectively.}\label{figS1}
\end{figure}

\begin{figure}[h]
	\centering
	\includegraphics[width=0.99\linewidth]{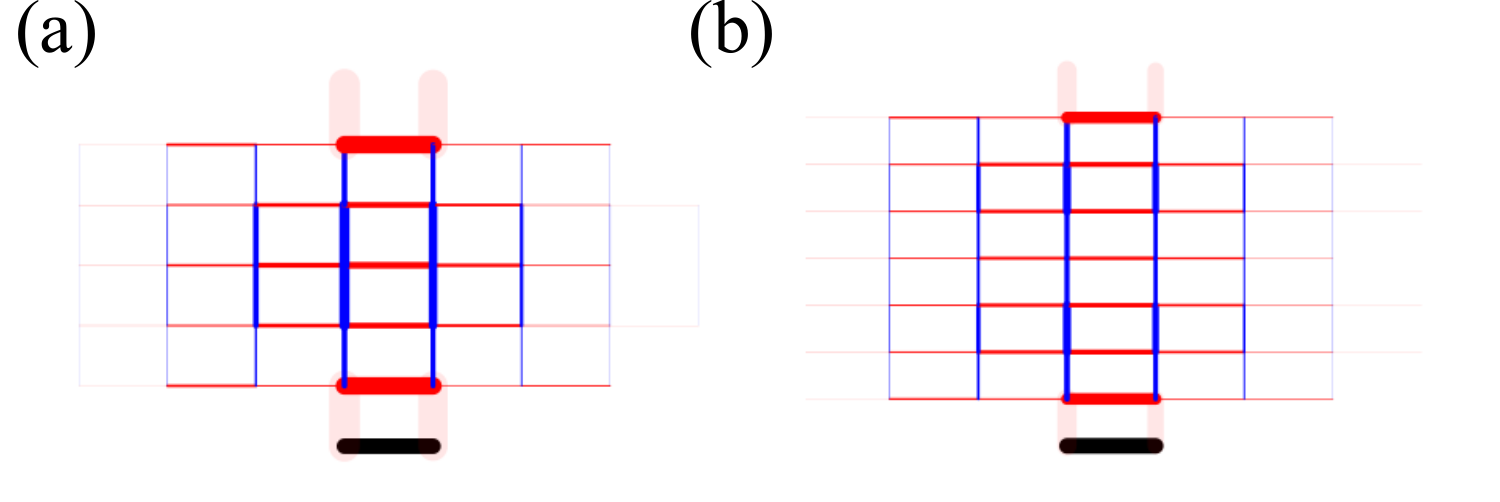}
	\caption{The PPC function, starting from a target $x$-bond (black line), is calculated using QCSM in the $t$-$J$ model. The parameters in (a,b) are the same as those used in Fig.~3(b) and Fig.~1(b) of the main text, respectively.
	}\label{figS2}
\end{figure}

The $d$-wave pairing pattern means that the PPCs between $x$-bonds and $y$-bonds are negative, while those between $x$-bonds are positive. To complement the data shown in the main text, we present the PPC functions of the $t$-$t'$-$U$ and $t$-$J$ models starting from a target $x$-bond.
As shown in Fig.~\ref{figS1} and Fig.~\ref{figS2}, the PPC functions obtained from both DMRG and QCSM exhibit similar $d$-wave patterns.

\begin{figure}[h]
	\centering
	\includegraphics[width=0.99\linewidth]{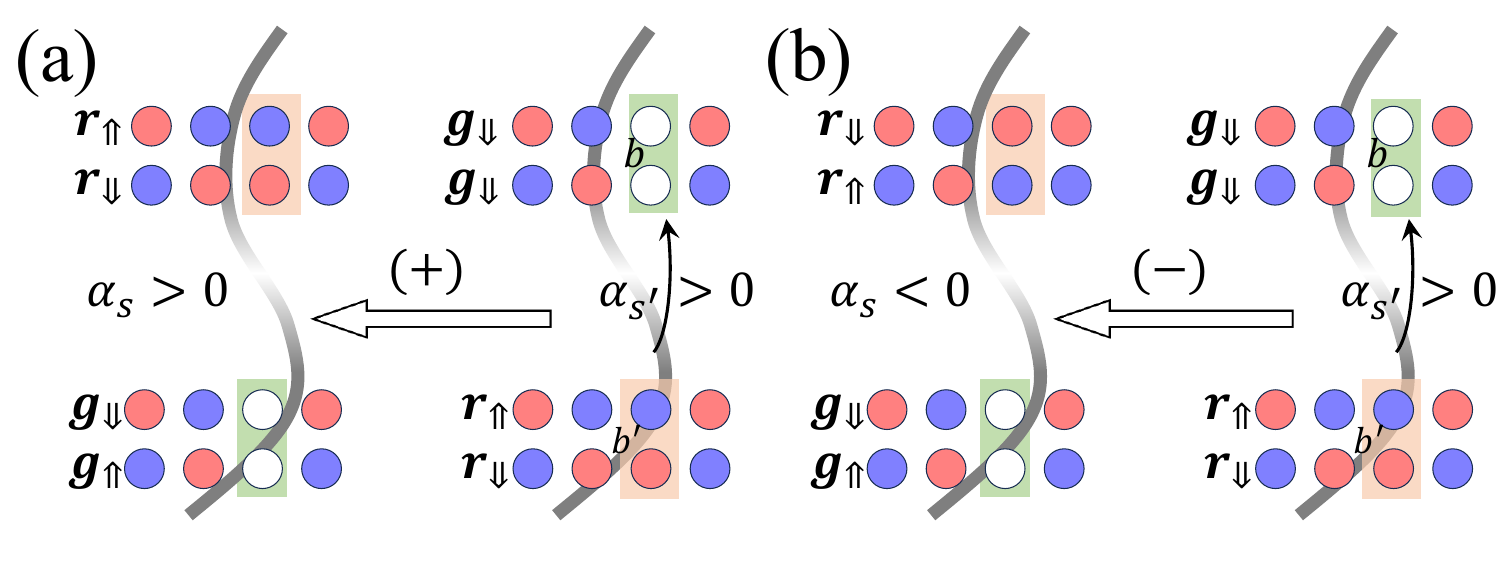}
	\caption{Schematic diagrams illustrate the processes of moving a pair of electrons from $y$-bond $b'$ to $y$-bond $b$ (a) with and (b) without changing the chirality sequence.
	}\label{figS4}
\end{figure}

\section{Microscopic Mechanism of $d$-wave}
In this section, we will discuss the underlying relation between spinon singlets and $d$-wave pairing pattern. 

\textit{\textbf{$y$-$y$ bonds}}.--As explained in the main text, the PPC functions between two $y$-bonds are mainly contributed by the two-spinon bases $\ket{s}$ (left panel) and $\ket{s'}$ (right panel) demonstrated in Fig.~\ref{figS4}.
The chirality order of the spinon pair will change if two electrons move to the positions of two holes, with a corresponding spin exchange.
Therefore, the overall sign of the PPC function will always be positive. 

\begin{figure}[h]
	\centering
	\includegraphics[width=0.99\linewidth]{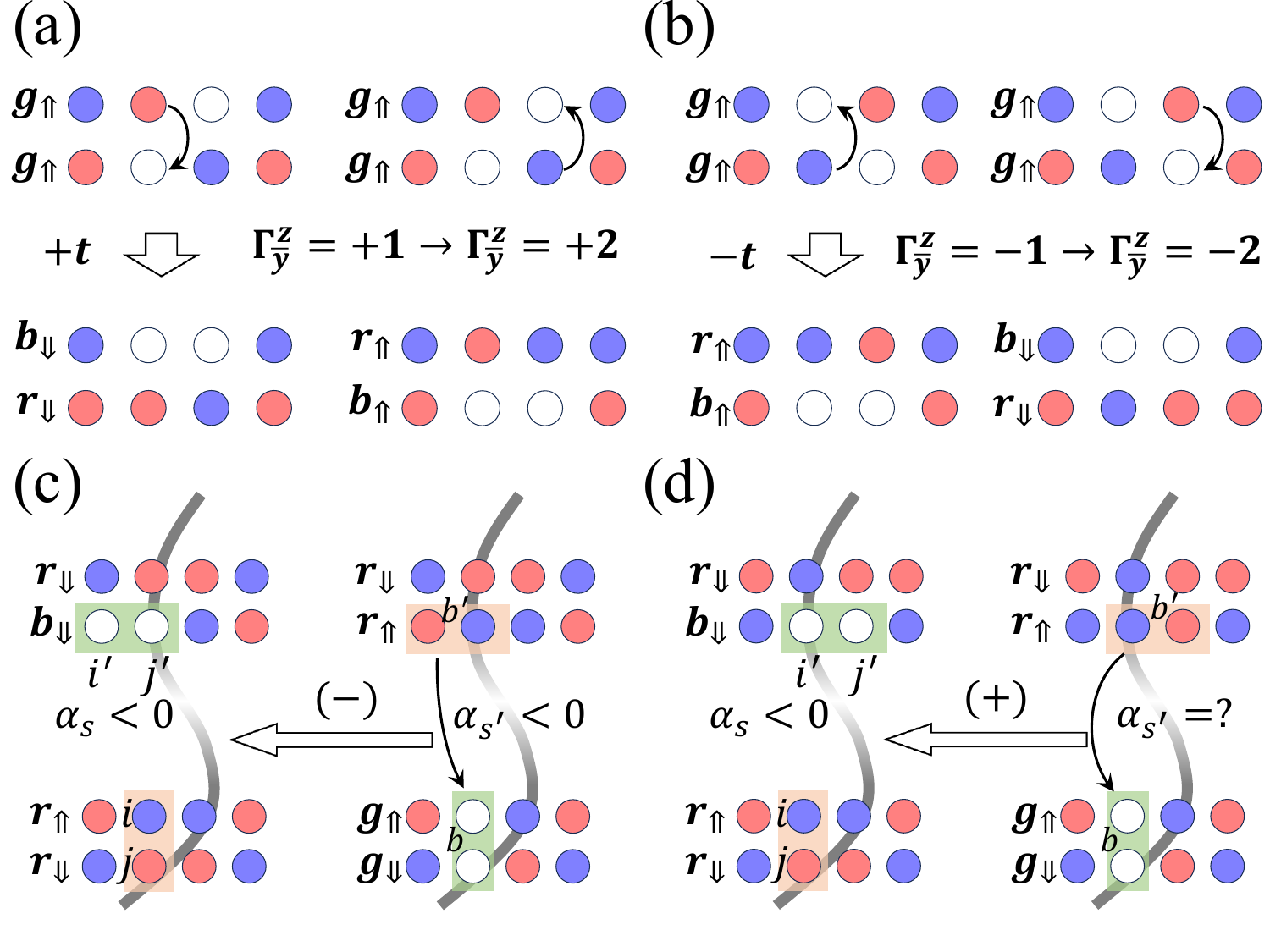}
	\caption{Schematic diagrams for (a,b) $(\textbf{r}\text{-}\textbf{b} / \textbf{g}\text{-}\textbf{g})$ processes which gives different signs, and (c,d) illustrate the processes of moving a electron pair from $x$-bond $b'$ to $y$-bond $b$. The negative sign of $\alpha_s$ result from the local fluctuation $\textbf{r}\text{-}\textbf{b}$ pair with positive effective spin field $\Gamma^z_{\bar{y}}=+2$.
	}\label{figS5}
\end{figure}

\textit{\textbf{$x$-$y$ bonds}}.--The analysis in the main text lacks a discussion of $\textrm{Sgn}_0$.
As shown in Figs.~\ref{figS5}(a,b), the sign of $(\textbf{r}\text{-}\textbf{b} / \textbf{g}\text{-}\textbf{g})$ process is closely related to the sign of the effective spin field between the holons~\cite{color_string}.
However, by comparing Fig.~\ref{figS5}(c) with Fig.~4(c) of the main text, we find that although $\textrm{Sgn}_0$ introduces an additional negative sign in $\alpha_s$, $\textrm{Sgn}_{\Delta}$ is also reversed simultaneously.
Furthermore, the process shown in Fig.~\ref{figS5}(d) can also contribute to the PPC function, but the sign of $\alpha_{s'}$ is dependent on the strength of the spin exchange interaction.
Fortunately, the amplitude of $\alpha_{s'}$ is very small because two closely spaced spinons with large $\Gamma_{\bar{y}}^z$ experience higher tension energy.

\begin{figure}[t!]
	\centering
	\includegraphics[width=0.99\linewidth]{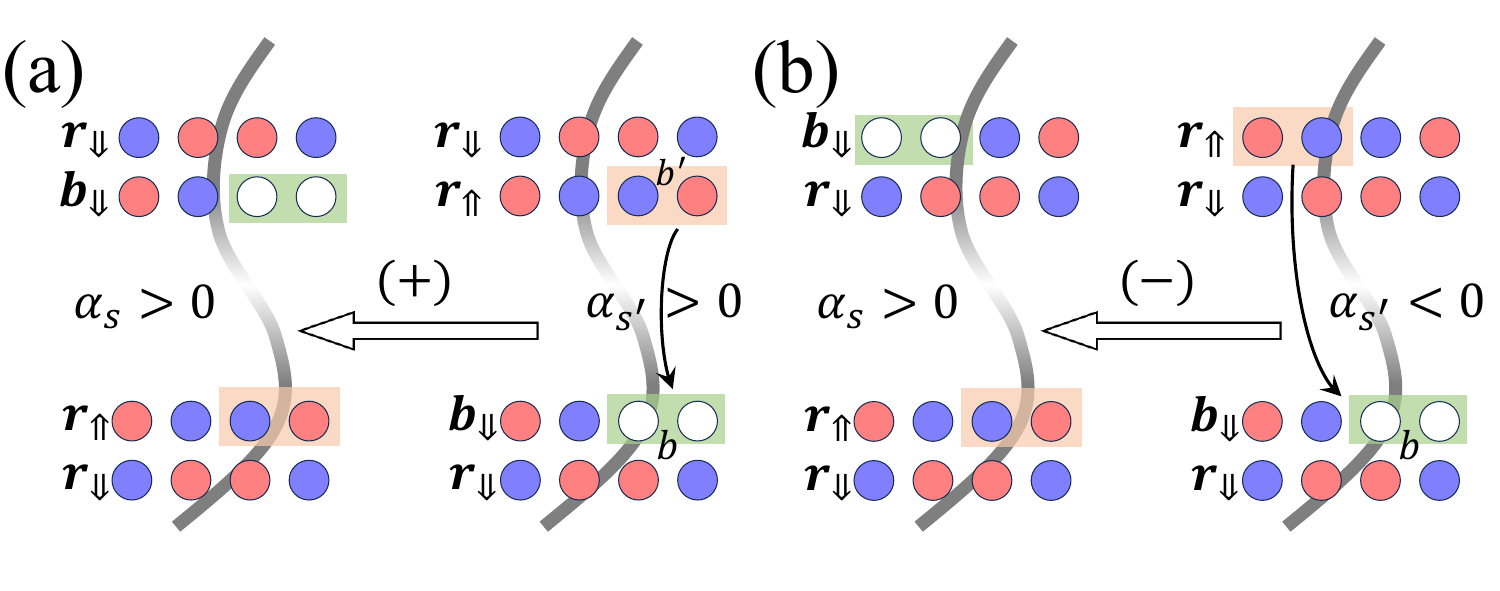}
	\caption{Schematic diagrams illustrate the processes of moving a electron pair from $x$-bond $b'$ to $x$-bond $b$ (a) with and (b) without changing the chirality sequence.
	}\label{figS6}
\end{figure}

\textit{\textbf{$x$-$x$ bonds}}.--The corresponding mechanism can also be understood using the QCSM framework established in the main text.
Unlike other cases, the main contributions to the PPC between two $x$-bonds stem from the movement of an electron pair along the $x$-axis, involving two three-spinon bases, $\ket{s}$ and $\ket{s'}$, as depicted in Fig.~\ref{figS6}.
When a spin exchange operation occurs between $\Delta_b$ and $\Delta_{b'}$, the chirality sequence is altered.
As a result, the sign of $G^{s,s'}_{b,b'}$ remains positive as long as the spinon-singlet is far from the \textbf{r}-\textbf{b} pair.
Furthermore, other processes related to the PPC between $x$-bonds can also be analyzed similarly, with the overall conclusion remaining the same.

At last, the amplitude of a two-spinon basis is much larger than that of a three-spinon basis when $J$ is not too small.
So the strengths of the PPC function follow the general relation: $x$-$x$ bonds $<$ $x$-$y$ bonds
$<$ $y$-$y$ bonds. 
{This implies that the $d$-wave is not isotropic in space, which may be attributed to the thin cylinders studied in the main text. However, since the local vacuum fluctuations, responsible for producing the \textbf{r}-\textbf{b} pairs, originates from electron hopping processes}, the difference in the PPC function between the $x$ and $y$ directions is expected to become smaller when $t/J$ increases.

\section{Pinning field in the DMRG calculations}
\begin{figure}[t]
	\centering
	\includegraphics[width=0.99\linewidth]{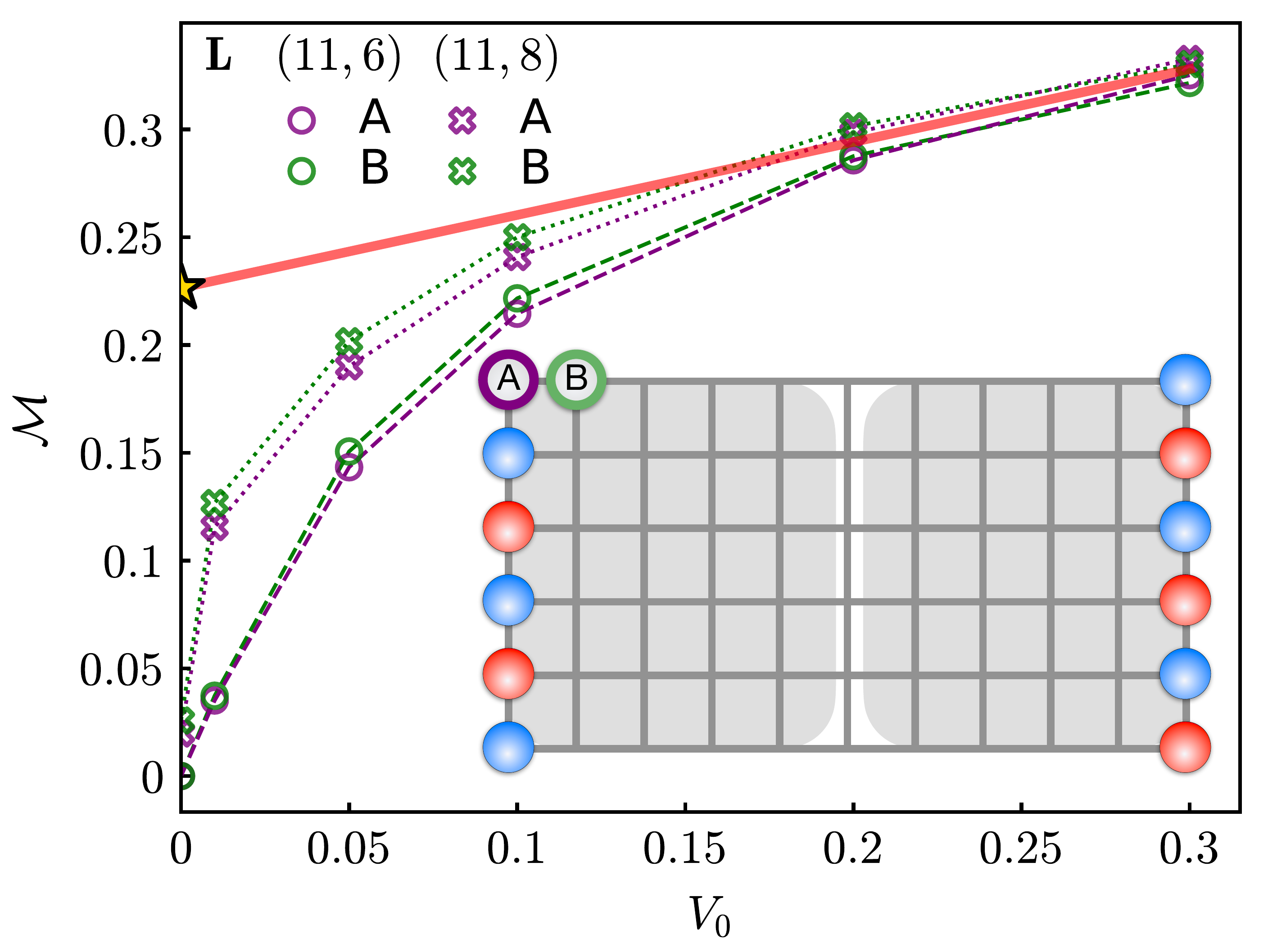}
	\caption{Local magnetizations $\mathcal{M} = \braket{S^z_\ell} = \braket{n_{\ell,\uparrow} - n_{\ell,\downarrow}}$, for two specific lattice sites A (violet) and B (green), as a function of the pinning field magnitude $V_0$.
		Specifically, a magnetic field along the positive and negative directions of the $z$-axis is applied to the blue and red sites, respectively, at two edge columns of the $t$-$J$ model.
		For a finely-tuned $V_0$, the orders in the left and right AF domains (grey shaded regions) are pinned along the $z$-axis.
		The other parameters are the same as those used in Fig.~1(b) and Fig.~3(d) of the main text, respectively.
	}\label{figS9-1}
\end{figure}

In the $t$-$J$ model, to facilitate perfect sampling, we pin the AF order in the left and right domains along the specific $z$-axis, {ensuring that the local spin fluctuations within AF domains do not contribute to the statistical data for spinons in QCSs}. This operation breaks the spin rotational symmetry in both domains, which is protected by finite energy gaps between quasi-degenerate energy levels.
Specifically, the pinning field adopts a staggered form, as shown in the inset of Fig.~\ref{figS9-1}, applying a magnetic filed along the $z$-axis with a magnitude of $\pm V_0$ to the red and blue sites, respectively.

In Fig.~\ref{figS9-1}, we note that when the cylinder circumference is $L_y=6$, the local magnetizations increase rapidly for a small pinning field magnitude $V_0$, and this behavior becomes much more pronounced as the cylinder circumference increases, for example, when $L_y=8$.
This phenomenon indicates spontaneous magnetization, which would arise in the thermodynamical limit as both $L_x$ and $L_y$ approach infinity.
As $V_0$ becomes large, the magnetization approaches a linear function of $V_0$ (red line), with a constant slope corresponding to the antiferromagnetic susceptibility.
The line has a finite intercept with the $y$-axis, revealing a finite antiferromagnetic moment due to spontaneous magnetization at $V_0=0$.
Notably, the system enters the symmetry-breaking region when $V_0\gtrsim 0.1$.

\begin{figure}[b]
	\centering
	\includegraphics[width=0.9\linewidth]{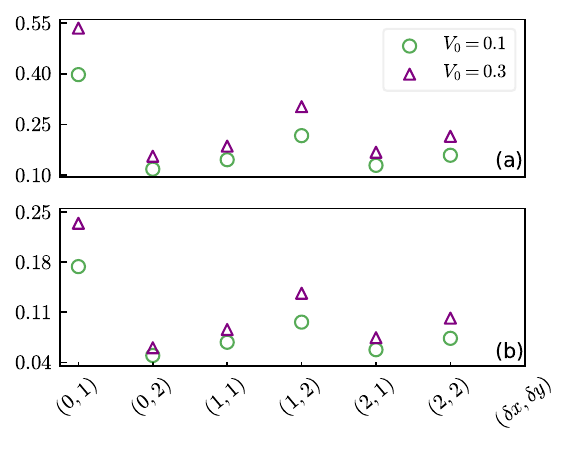}
	\caption{Spinon correlation function with representative spinon distances $(\delta x, \delta y)$ for the $t$-$J$ model. Here we make a comparison between $V_0=0.1$ and $0.3$ when fixing the truncated bond dimension of $M=8,192$.
		The other parameters in the plots (a,b) are the same as those used in Fig.~1(b) and Fig.~3(d) of the main text, respectively.
	}\label{figS9-2}
\end{figure}

We further show the benchmark for sampling data using two typical pinning field magnitudes in the symmetry-breaking region $V_0=0.1$ and $0.3$.
In Fig.~\ref{figS9-2}, we observe that the spinon correlation function for representative spinon distances consistently increases as $V_0$ increases, without changing the ordering of the spinon correlation values.
Meanwhile, the snapshots for the spinon singlet with $(\delta x, \delta y) = (0, 1)$ dominate.
Therefore, in the main text, we always use $V_0=0.3$ as the representative magnitude of the pinning field during the perfect sampling procedure.
The ratios of count and weight for the snapshots with spinon singlets depending on the pinning field magnitude are listed in Table~\ref{cwratiov0}.

\begin{table}[t]
	\setlength{\tabcolsep}{8pt}
	\centering
	\begin{tabular}{|c|c|c|c|c|}
		\toprule[1.5pt]
		\multicolumn{5}{|c|}{$t$-$J$ model at $J=0.6$}\\
		\midrule[1.5pt]
		\multicolumn{5}{|c|}{A $1/2$ hole-filled stripe with $\mathbf{L}=(11,\, 8)$}\\
		\hline
		& \multicolumn{2}{c|}{count ($\%$)} & \multicolumn{2}{c|}{weight ($\%$)}\\
		\hline
		$V_0$ & $0.1$ & $0.3$ & $0.1$ & $0.3$ \\
		\hline
		ratio & $26.2$ & $42.4$  & $93.0$ & $96.2$ \\
		\midrule[1.5pt]
		\multicolumn{5}{|c|}{A $2/3$ hole-filled stripe with $\mathbf{L}=(11,\, 6)$}\\
		\hline
		& \multicolumn{2}{c|}{count ($\%$)} & \multicolumn{2}{c|}{weight ($\%$)}\\
		\hline
		$V_0$ & $0.1$ & $0.3$ & $0.1$ & $0.3$ \\
		\hline
		ratio & $15.7$ & $21.2$  & $56.0$ & $55.6$ \\
		\bottomrule[1.5pt]
	\end{tabular}
	\caption{The ratios of count and weight for the snapshots with spinon singlet pairs as a function of the pinning field magnitude $V_0$, when $M=8,192$ is used in DMRG.}
	\label{cwratiov0}
\end{table}

\section{Systematic errors in the DMRG calculations}
In Table~\ref{errors}, we show the truncation errors for three distinct sets of hole filling in the stripe, cylinder circumference $L_y$, and other parameters in the $t$-$t'$-$U$ and $t$-$J$ models studied in this work.
We find that a truncated bond dimension of $M=16,384$ ensures that the truncation errors are below or on the order of $\lesssim 10^{-4}$, when the pinning field is absent. With a finite pinning field, the truncation errors become consistently smaller compared to those for $V_0=0$.

\begin{table}[h]
	\centering
	\begin{tabular}{|c|c|c|c|c|c|}
		\toprule[1.5pt]
		\multicolumn{6}{|c|}{$t$-$t'$-$U$ model at $t_2=-0.2$ and $U=12$}\\
		\midrule[1.5pt]
		\multicolumn{6}{|c|}{A $2/3$ hole-filled stripe with $\mathbf{L}=(9,\, 6)$}\\
		\hline
		$M$ & $2,048$ & $4,096$ & $8,192$ & $12,288$ & $16,384$ \\
		\hline
		$\epsilon$ & \quad\quad $\slash$ \quad\quad\quad & \quad\quad $\slash$ \quad\quad\quad & $1.19\times10^{-4}$ & $9.99\times10^{-5}$  & $9.00\times10^{-5}$
		\\
		\bottomrule[1.5pt]
	\end{tabular}
	\begin{tabular}{|c|c|c|c|c|c|}
		\toprule[1.5pt]
		\multicolumn{6}{|c|}{$t$-$J$ model at $J=0.6$}\\
		\midrule[1.5pt]
		\multicolumn{6}{|c|}{A $2/3$ hole-filled stripe with $\mathbf{L}=(11,\, 6)$}\\
		\hline
		$M$ & $2,048$ & $4,096$ & $8,192$ & $12,288$ & $16,384$ \\
		\hline
		$\epsilon$ & $2.31\times10^{-4}$ & $9.12\times10^{-5}$ & $3.15\times10^{-5}$ & $1.50 \times 10^{-5}$
		& $8.47 \times 10^{-6}$
		\\
		\midrule[1.5pt]
		\multicolumn{6}{|c|}{A $1/2$ hole-filled stripe with $\mathbf{L}=(11,\, 8)$} \\
		\hline
		$M$ & $2,048$ & $4,096$ & $8,192$ & $12,288$ & $16,384$ \\
		\hline
		$\epsilon$ & $4.12\times10^{-4}$ & $2.95\times10^{-4}$ & $1.96\times10^{-4}$ & $1.28\times 10^{-4}$
		& $9.19\times 10^{-5}$
		\\
		\bottomrule[1.5pt]
	\end{tabular}
	\caption{The truncation errors $\epsilon$ as a function of the truncated bond dimension $M$ used in DMRG, when the pinning field is absent.}
	\label{errors}
\end{table}

\begin{figure}[t]
	\centering
	\includegraphics[width=0.88\linewidth]{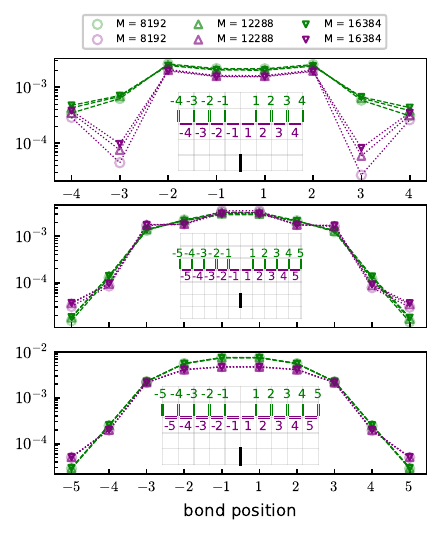}
	\caption{The PPC function for $y$-$y$ bonds (green symbols) and $y$-$x$ bonds (violet symbols) as a function of the bond position for (a) the $t$-$t'$-$U$ model, and (b,c) the $t$-$J$ model.
		Insets: the PPC function, starting from a target $y$-bond (black line),  is measured along a path of $y$-bonds (green lines) and another path of $x$-bonds (violet lines) for a comparison. Numbers mark the bond positions. 
		The other parameters are the same as those used in Figs.~1(a,b) and Fig.~3(d) of the main text, respectively.
	}\label{figS7-1}
\end{figure}

\begin{table}[t]
	\setlength{\tabcolsep}{5.5pt}
	\centering
	\begin{tabular}{|c|c|c|c|c|c|c|}
		\toprule[1.5pt]
		\multicolumn{7}{|c|}{$t$-$t'$-$U$ model at $t_2=-0.2$ and $U=12$}\\
		\midrule[1.5pt]
		\multicolumn{7}{|c|}{A $2/3$ hole-filled stripe with $\mathbf{L}=(9,\, 6)$}\\
		\hline
		& \multicolumn{3}{c|}{count ($\%$)} & \multicolumn{3}{c|}{weight ($\%$)}\\
		\hline
		$M$ & $8,192$ & $12,288$ & $16,384$ & $8,192$ & $12,288$ & $16,384$ \\
		\hline
		ratio & $13.8$ & $13.3$  & $12.8$ & $38.5$ & $38.1$ & $41.8$
		\\
		\bottomrule[1.5pt]
	\end{tabular}
	\begin{tabular}{|c|c|c|c|c|c|c|}
		\toprule[1.5pt]
		\multicolumn{7}{|c|}{$t$-$J$ model at $J=0.6$}\\
		\midrule[1.5pt]
		\multicolumn{7}{|c|}{A $1/2$ hole-filled stripe with $\mathbf{L}=(11,\, 8)$}\\
		\hline
		& \multicolumn{3}{c|}{count ($\%$)} & \multicolumn{3}{c|}{weight ($\%$)}\\
		\hline
		$M$ & $8,192$ & $12,288$ & $16,384$ & $8,192$ & $12,288$ & $16,384$ \\
		\hline
		ratio & $42.4$ & $42.0$  & $42.0$ & $96.2$ & $96.1$ & $96.0$
		\\
		\midrule[1.5pt]
		\multicolumn{7}{|c|}{A $2/3$ hole-filled stripe with $\mathbf{L}=(11,\, 6)$}\\
		\hline
		& \multicolumn{3}{c|}{count ($\%$)} & \multicolumn{3}{c|}{weight ($\%$)}\\
		\hline
		$M$ & $8,192$ & $12,288$ & $16,384$ & $8,192$ & $12,288$ & $16,384$ \\
		\hline
		ratio & $21.2$ & $21.3$  & $21.2$ & $55.6$ & $55.7$ & $56.1$
		\\
		\bottomrule[1.5pt]
	\end{tabular}
	\caption{The ratios of count and weight for the snapshots with spinon singlet pairs as a function of the truncated bond dimension $M$ used in DMRG, when $V_0=0.3$.}
	\label{cwratio}
\end{table}

We further benchmark the PPC function, starting from a target $y$-bond, along two given paths shown in the insets of Fig.~\ref{figS7-1}.
By varying the truncated bond dimensions from $M=8,192$ to $16,384$ across three typical parameter sets, we find that the absolute values of the PPC function remain almost unchanged across a broad range, spanning two orders of magnitudes, e.g., from $10^{-2}$ to $10^{-5}$ as shown in Fig.~\ref{figS7-1}(a).
Therefore, the PPC function data presented in the main text is quantitatively reliable for demonstrating electron pairing.

\begin{figure}[b]
	\centering
	\includegraphics[width=0.9\linewidth]{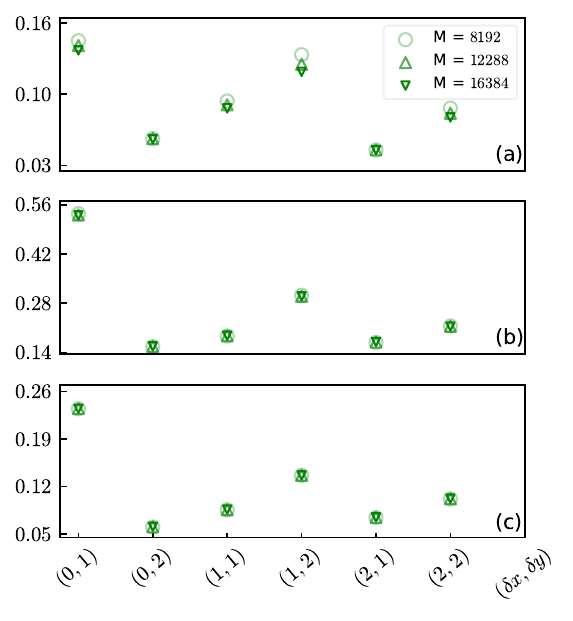}
	\caption{Spinon correlation function with spinon distances $(\delta x, \delta y)$ for (a) the $t$-$t'$-$U$ model, and (b,c) the $t$-$J$ model.
		The pinning field magnitude $V_0=0.3$ is used in the perfect sampling procedure.
		The other parameters are the same as those used in Figs.~1(a,b) and Fig.~3(d) of the main text, respectively.
	}\label{figS7-2}
\end{figure}

In Fig.~\ref{figS7-2}, we also present a benchmark of the statistical data obtained from the perfect sampling technique.
As the truncated bond dimension $M$ increases, we find that the spinon correlation function for representative spinon distances exhibit small errors well below the symbol sizes, for two specific parameter sets in the $t$-$J$ model, as show in Figs.~\ref{figS7-2}(b,c).
In contrast, the $t$-$t'$-$U$ model shows a more noticeable change as $M$ increases in Fig.~\ref{figS7-2}(a).
For the case of spinon distance $(1, 2)$, the relative change is roughly less than $15\%$ when comparing the data for $M=8,192$ and $16,384$.
Nevertheless, the snapshots for the spinon singlet with spinon distance $(0,1)$ cintinue to dominate the ground-state wave functions, which is a key feature for determining the $d$-wave pairing pattern for electrons, as discussed in the main text.
The ratios of count and weight for the snapshots with spinon singlets versus the truncated bond dimensions are listed in Table~\ref{cwratio}.

\section{Details of ``unpaired" and ``invalid"}
\begin{figure}[h]
	\centering
	\includegraphics[width=0.99\linewidth]{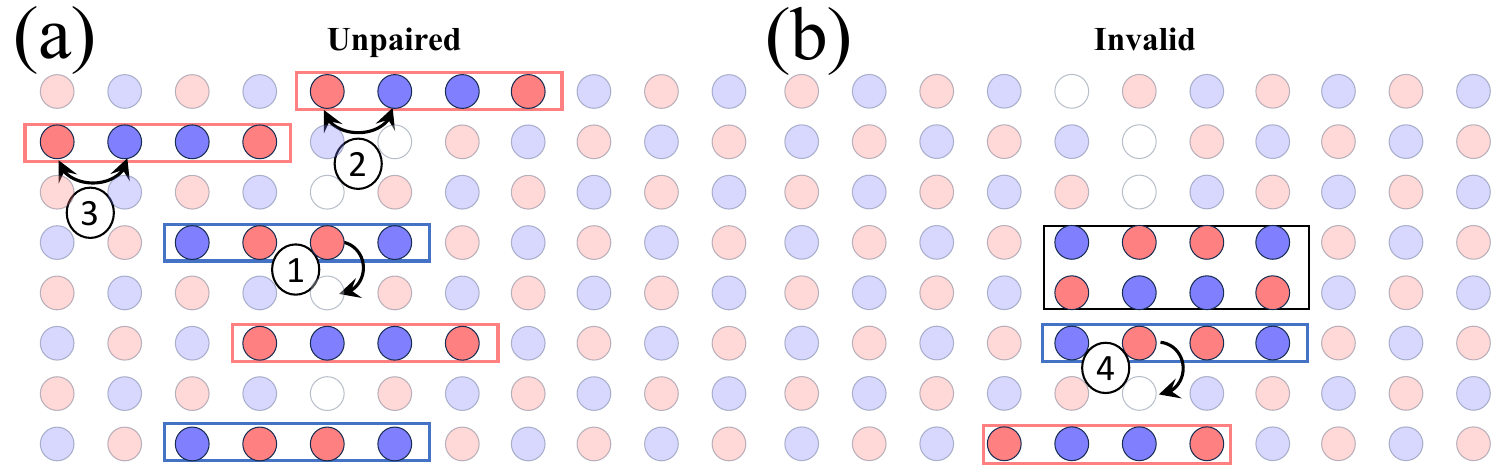}
	\caption{The typical snapshots of (a) ``unpaired" and (b) ``invalid" cases for a $1/2$ hole-doped stripe in the $t$-$J$ model. The black arrows with numbers in circles label different dynamic processes.
		All the data are obtained via the perfect sampling technique, and the parameters are the same as those used in Fig.~1.
	}\label{figS8}
\end{figure}

The appearance of ``unpaired" and ``invalid" samples can be understood by analyzing the typical snapshots presented in Figs.~\ref{figS8}(a,b), respectively. Through the hopping process~\process{1} and spin-exchange process~\process{2}, the {spinon pair} can be restored.
It is clear that these two consecutive processes enable the spinon pair to move in the $y$ and $x$ directions, respectively. Meanwhile, the quantum fluctuation process~\process{3} can locally destroy the AF order, but such excitation is unstable due to its high energy cost. Therefore, ``unpaired" samples mainly arise from local perturbations of ``spinon singlet" states, which explains why many ``unpaired" samples have apparently much smaller weights.
The process~\process{4} in Fig.~2(b) demonstrates that one spinon in a spinon singlet can meet another spinon singlet. However, it cannot break up two existing spinon singlets to form a new spinon singlet, which is the main reason for the ``invalid" part.

\section{Benchmarks of model parameters}
The $d$-wave pattern of the PPC function can be affected by factors, such as system size, doping concentration, and other tunable parameters defined in the model Hamiltonian.
In this section, we examine the $t$-$J$ model and the $t$-$t'$-$U$ model with varying model parameters, and all calculations are performed using DMRG with $M=8,192$.

\begin{figure}[b]
	\centering
	\includegraphics[width=0.99\linewidth]{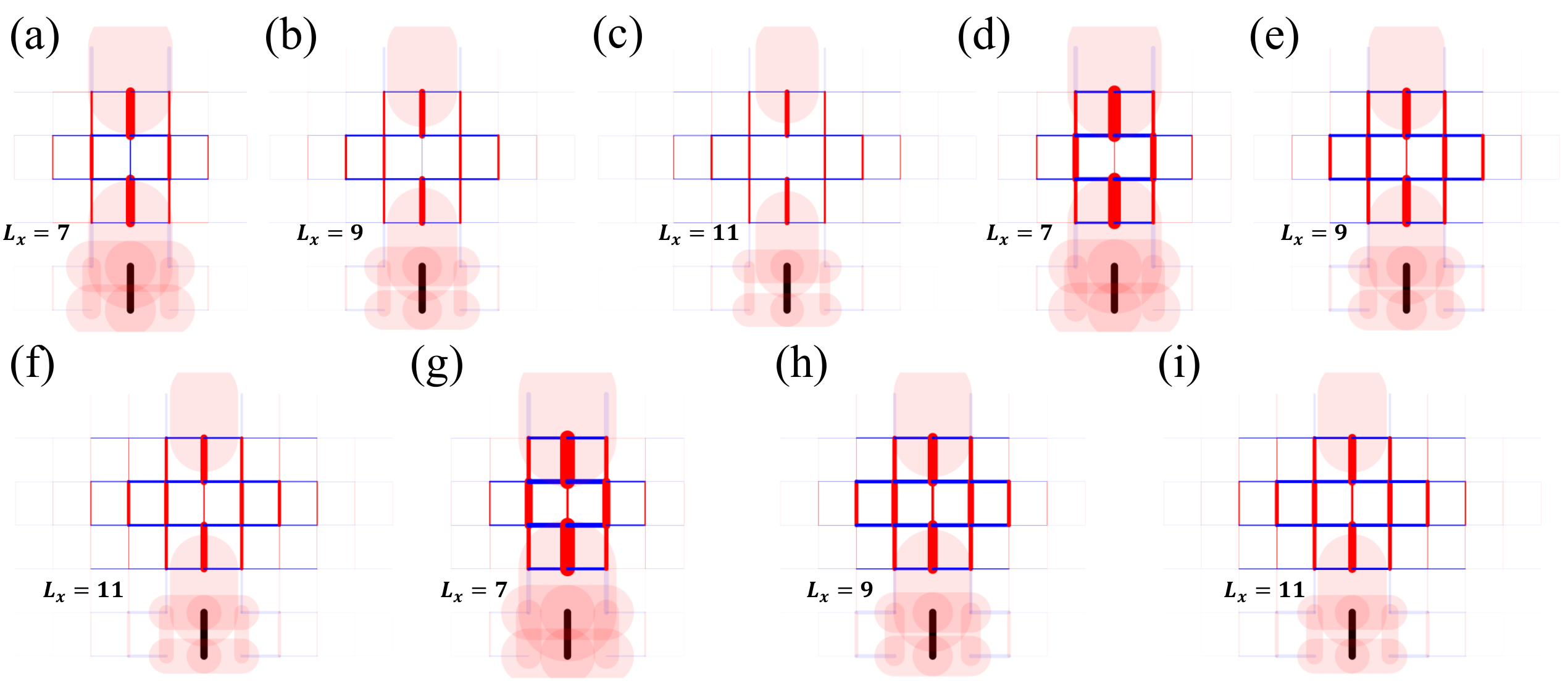}
	\caption{The PPC function for a $2/3$ hole-doped stripe, starting from a target $y$-bond (black line), in the $t$-$J$ model at (a-c) $J=1/3$, (d-f) $0.5$, and (g-i) $0.6$, with $L_y=6$ and varying $L_x$.
	}\label{figS9}
\end{figure}

First, as demonstrated in Fig.~\ref{figS9}, we change $J$ and $L_x$ in the $t$-$J$ model while keeping $L_y=6$ fixed, and observe that all the PPC functions exhibit a clear $d$-wave pattern.
Furthermore, we find that a larger spin interaction $J$ strengthens the $d$-wave correlation at long distances.
Meanwhile, the weight of snapshots for the spinon singlet, as listed in Table~\ref{table_w01}, increases, further confirming that the spinon singlet pairs serve as carriers of the $d$-wave superconducting correlation.
The dependence of these ratios on $J$ is found to be similar to that in the $t$-$t'$-$U$ model with $L_y=6$, with the ratios increasing as $J = 4 t^2 / U$ grows, or equivalently, as $U$ decreases.

\begin{table}[h]
	\setlength{\tabcolsep}{13pt}
	\centering
	\begin{tabular}{|c|c|c|c|}
		\toprule[1.5pt]
		\multicolumn{4}{|c|}{$t$-$J$ model: a $2/3$ hole-doped stripe with $L_y=6$ }\\
		\midrule[1.5pt]
		& $L_x=7$ & $L_x=9$ & $L_x=11$ \\
		\hline
		$J=1/3$ & $34.8$ & $45.3$ & $48.5$ \\
		\hline
		$J=0.5$ & $39.8$ & $51.4$ & $52.6$ \\
		\hline
		$J=0.6$ & $42.3$ & $55.3$ & $56.4$ \\
		\midrule[1.5pt]
		\multicolumn{4}{|c|}{$t$-$J$ model: a $1/2$ hole-doped stripe with $L_y=8$ }\\
		\midrule[1.5pt]
		& $L_x=5$ & $L_x=7$ & $L_x=9$ \\
		\hline
		$J=1/3$ & $43.1$ & $85.0$ & $92.5$ \\
		\hline
		$J=0.5$ & $41.0$ & $88.6$ & $96.2$ \\
		\hline
		$J=0.6$ & $40.0$ & $89.4$ & $96.0$ \\
		\midrule[1.5pt]
		\multicolumn{4}{|c|}{$t$-$t'$-$U$ model: a $2/3$ hole-doped stripe with $L_y=6$ }\\
		\midrule[1.5pt]
		& $L_x=7$ & $L_x=9$ & $L_x=11$ \\
		\hline
		$U=8$ & $30.6$ & $29.5$ & $17.6$ \\
		\bottomrule[1.5pt]
	\end{tabular}
	\caption{The ratios of weight for the snapshots with spinon singlet pairs marked in percentage points ($\%$).}
	\label{table_w01}
\end{table}

\begin{figure}[h]
	\centering
	\includegraphics[width=0.99\linewidth]{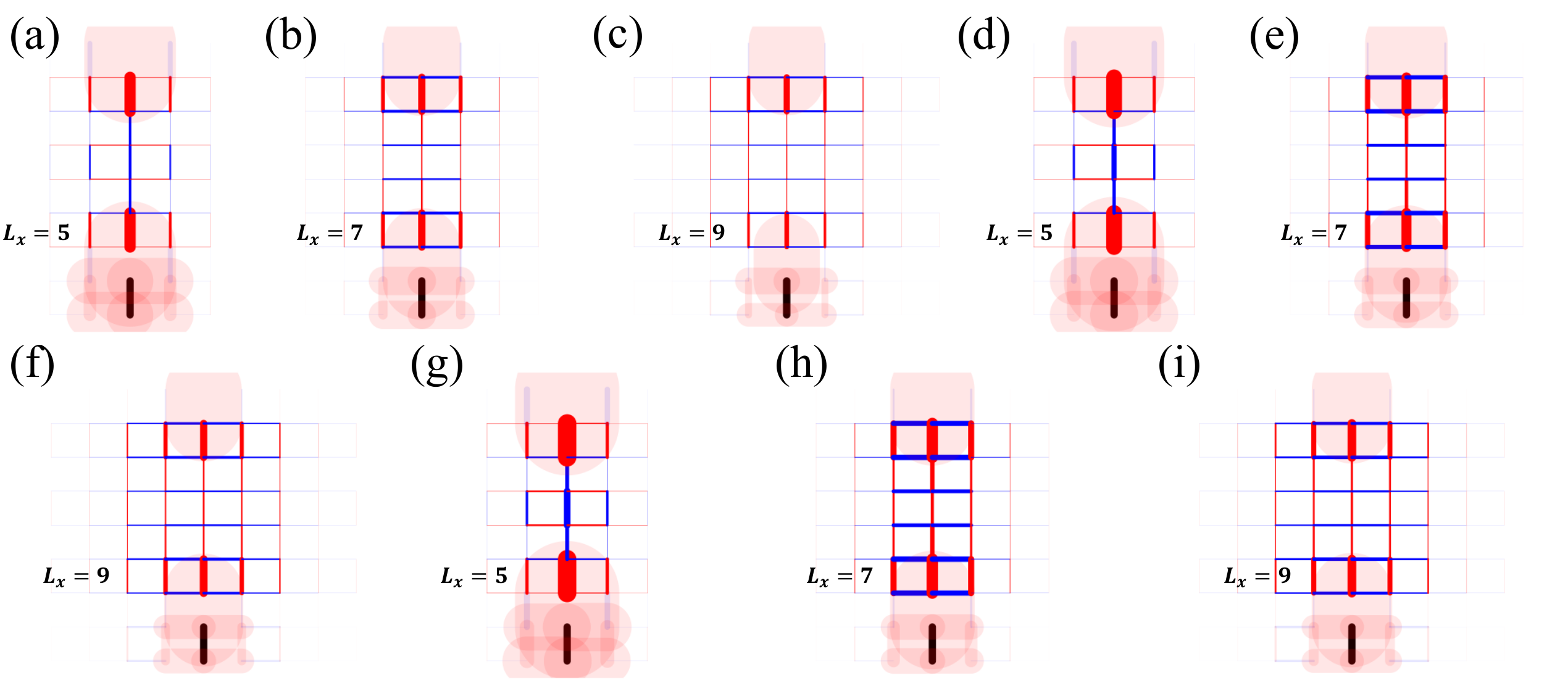}
	\caption{The PPC function for a $1/2$ hole-doped stripe, starting from a target $y$-bond (black line), in the $t$-$J$ model at (a-c) $J=1/3$, (d-f) $0.5$, and (g-i) $0.6$, with $L_y=8$ and varying $L_x$.
	}\label{figS10}
\end{figure}

\begin{figure}[t]
	\centering
	\includegraphics[width=0.99\linewidth]{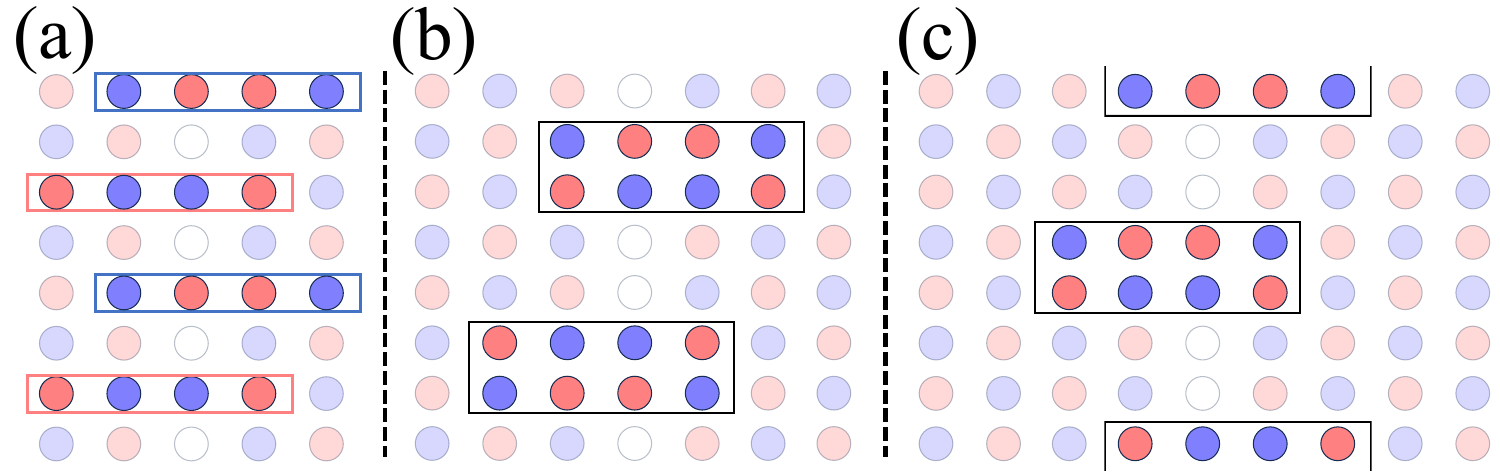}
	\caption{The typical snapshots for a $1/2$ hole-doped stripe in the $t$-$J$ model at $J=0.6$ with (a) $L_x=5$, (b) $7$, and (c) $9$. 
		All the data are obtained via the perfect sampling technique.
	}\label{figS11}
\end{figure}

For a $1/2$ hole-doped stripe, as shown in Fig.~\ref{figS10}, the $d$-wave patterns remain robust for different values of $J$ with $L_x=7$ and $9$.
However, for $L_x=5$, all the PPC functions present a different sign structure [Fig.~\ref{figS10}(a,d,g)], rather than the uniform $d$-wave patterns.
Such change is also reflected in The ratios of weight for the snapshots with spinon singlet pairs, as listed in Table~\ref{table_w01}, in which a sudden drop is observed when $L_x$ decreases from $7$ to $5$.
To understand the underlying physical reason, we check the typical snapshots.
As shown in Fig.\ref{figS11}, it is clear that the spinons do not tend to form pairs at $L_x=5$ [Fig.\ref{figS11}(a)] in stark contrast to two spinon singlet pairs observed at $L_x=7$ and $9$ [Fig.\ref{figS11}(b,c)].
At small $L_x$, the fluctuations of the stripe along the $x$-axis are suppressed, and the stripe remains nearly static~\cite{Chen_2024}.
We infer that the spinons prefer to form a liquid along QCS at small $L_x$, possibly due to the closure of the spinon gap in the \textit{static stripe}.
Given that the doping concentration is $1/10$, which is close to the $1/8$ magic filling, the disappearance of the $d$-wave pattern may provide some clues to the suppression of superconducting correlations in LSCO at $1/8$ doping~\tcr{\cite{hightc_review3}}.

\begin{figure}[t]
	\centering
	\includegraphics[width=0.99\linewidth]{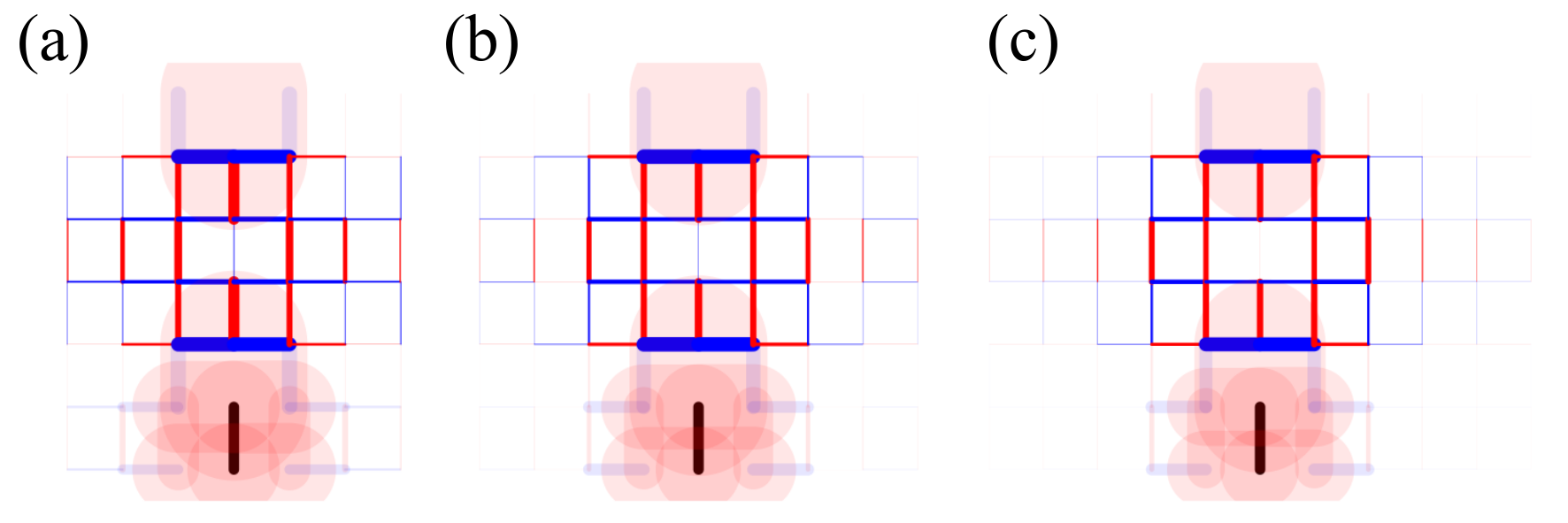}
	\caption{The PPC function for a $2/3$ hole-doped stripe, starting from a target $y$-bond (black line), in the Hubbard model including the next-nearest-neighboring (NNN) hopping $t'=-0.2$ and $U=8$ with $L_y=6$ and (a) $L_x=7$, (b) $L_x=9$, and (c) $L_x=11$. 
	}\label{figS12}
\end{figure}

\begin{figure}[b]
	\centering
	\includegraphics[width=0.99\linewidth]{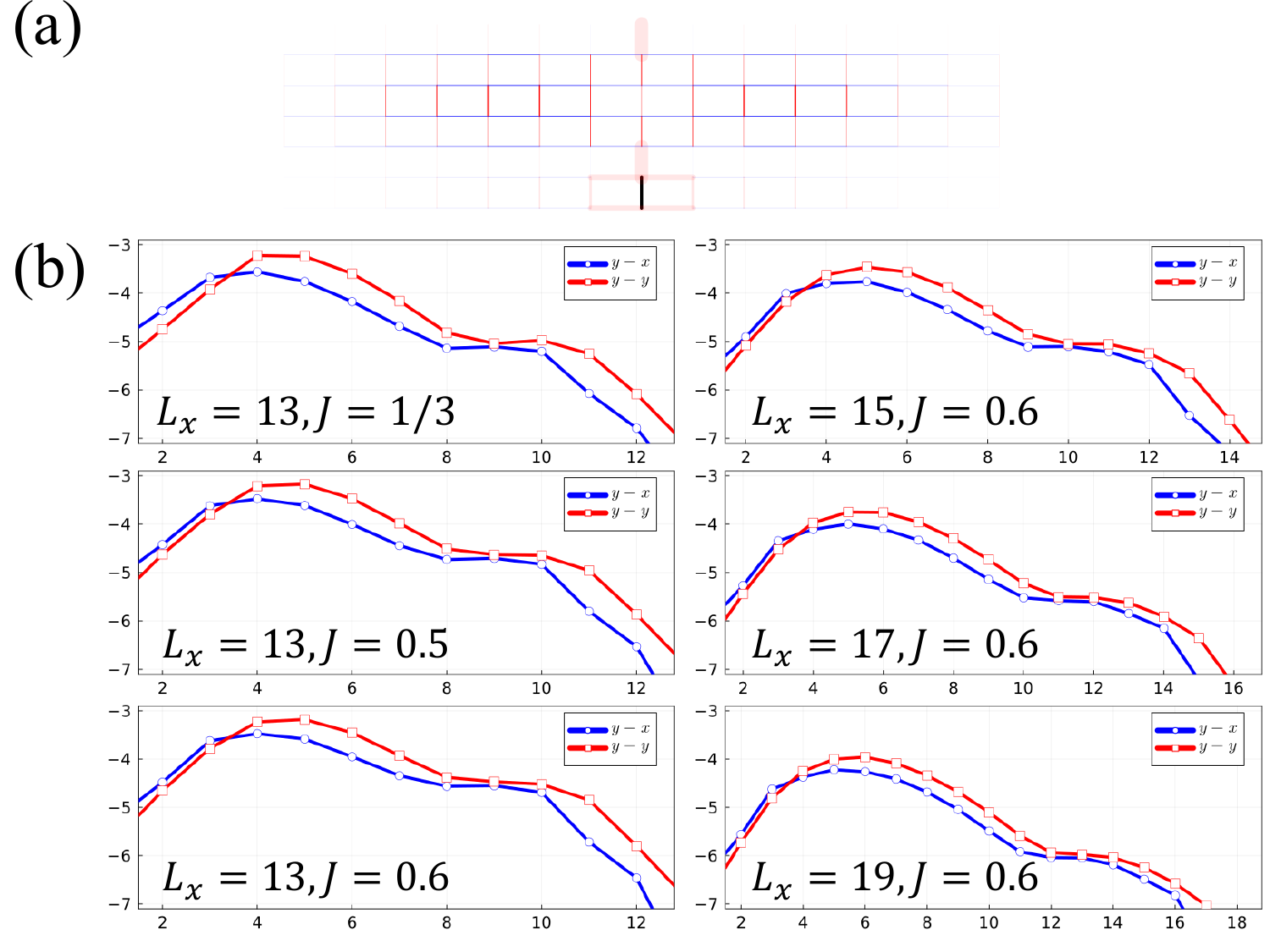}
	\caption{(a) The PPC function in the $t$-$J$ model with $N_h=8$ (two $2/3$ hole-doped stripes), and $\mathbf{L} = (19,\, 6)$, starting from a target $y$-bond (black line).
		(b) The PPC function $\mathrm{log}_{10}|G_{b,b'}|$ for different $J$ and $L_x$, where bonds $b$ and $b'$ are the same as those in Fig.~5(a) of the main text.
	}\label{figS13}
\end{figure}

Furthermore, in the Hubbard model with NNN hopping terms, we check the PPC function for smaller $U=8$ and $L_y=6$ at varying $L_x$.
As shown in Fig.~\ref{figS12}, the $d$-wave patterns are also stable.
However, the ratios of weight for the snapshots with spinon singlet pairs are relatively small, as listed in Table~\ref{table_w01}, compared to Fig.~2(d) of the main text for larger $U=12$.
This difference may be attributed to the larger higher-order processes at smaller $U$.

Finally, we benchmark the case of double stripes.
As shown in Fig.~\ref{figS13}(a), we observe that the PPC functions, starting from the target bond, still exhibit a strong $d$-wave pattern over a long distance, though the average distance between the two stripes increases when $L_x=19$, compared to Fig.~5(a) of the main text.
Moreover, in Fig.~\ref{figS13}(b), for different values of $J$ and $L_x$, we consistently find that the PPC function starting from $y$-bond $b$ near the left edge column is enhanced by the left QCS and further boosted by the right QCS over a long distance, which supports the conclusion for double stripes mentioned in the main text.
We also note that at $L_x=13$, the doping concentration is $0.103$, which is close to $1/8$.
However, the $d$-wave pattern remains robust.
Therefore, we argue that fluctuating stripes may enhance $d$-wave superconducting correlations more effectively than static stripes~\cite{Chen_2024}.
\end{document}